\documentclass[11pt]{article}
\usepackage{amssymb}
\usepackage{graphicx}
\usepackage{amsmath}
\usepackage{makeidx}
\usepackage{indentfirst}

\newcounter{resultnum}[section]
\newcounter{conclusionnum}[section]
\newcounter{conditionnum}[section]
\newcounter{conjecturenum}[section]
\newcounter{examplenum}[section]
\newcounter{exercisenum}[section]
\newcounter{lemmanum}[section]
\newcounter{notationnum}[section]
\newcounter{theoremnum}[section]
\newcounter{definitionnum}[section]
\newcounter{corollarynum}[section]
\newcounter{remarknum}[section]
\newcounter{propositionnum}[section]
\newcounter{acknowledgementnum}[section]
\newcounter{algorithmnum}[section]
\newcounter{axiomnum}[section]
\newcounter{casenum}[section]
\newcounter{claimnum}[section]
\newcounter{summarynum}[section]
\newcounter{problemnum}[section]
\begin{document}

\title{ Black Holes, Ellipsoids, and Nonlinear Waves in Pseudo--Finsler
Spaces and Einstein Gravity}
\date{December 8, 2012}
\author{ \textbf{Sergiu I. Vacaru}\thanks{%
sergiu.vacaru@uaic.ro,\ \ http://www.scribd.com/people/view/1455460-sergiu}
\and \textsl{\small Alexandru Ioan Cuza University, UAIC, } \\
\textsl{\small Alexandru Lapu\c sneanu street, nr. 14; Corpus R, office 323;
Ia\c si, Romania, 700057 } }
\maketitle

\begin{abstract}
We model pseudo--Finsler geometries, with pseudo--Euclidean signatures of
metrics, for two classes of four dimensional nonholonomic manifolds:\ a)
tangent bundles with two dimensional base manifolds and\ b)
pseudo--Riemannian/ Einstein spaces. Such spacetimes are enabled with
nonholonomic distributions and theirs metrics are solutions of the field
equations in general relativity and/or generalizations. We rewrite the
Schwarzschild metric in Finsler variables and use it for generating new
classes of black hole objects with stationary deformations to ellipsoidal
configurations. The conditions are analyzed when such metrics describe
imbedding of black hole solutions into nontrivial solitonic backgrounds.

\vskip0.1cm

\textbf{Keywords:}\ Pseudo--Finsler geometry, nonholonomic manifolds and
bundles, nonlinear connections, black holes and ellipsoids.

\vskip3pt 2000 MSC:\ 83C15, 83C57, 83C99, 53C60, 53B40

PACS:\ 04.20.Jb, 04.50.Kd, 04.70.Bw, 04.90.+e
\end{abstract}



\section{Introduction}

The goal of this work is to construct new classes of (Finsler) black hole
solutions and analyze their deformations to metrics with ellipsoidal
symmetry and (or) imbedding into nontrivial solitonic backgrounds. Such
models of pseudo--Finsler spacetimes can be elaborated on tangent bundles or
on pseudo--Riemannian manifolds enabled with the corresponding nonholonomic
frame structures. Let us motivate our interest in this problem:

Black holes are investigated in great depth and detail for more than fifty
years for all important gravity theories, for instance, in general
relativity and string/ brane gravity and their bimetric, gauge,
noncommutative modifications/ generalizations etc. We cite here monographs
\cite{haw,kramer,bic,heu} and a recent resource letter \cite{gallo} for
literature on black hole physics and mathematics. Various classes of exact
solutions describing generalized Finsler metrics were constructed in low and
extra dimensional gravity for spacetime models with nonholomomic
distributions, see reviews of results and methods in Refs. \cite%
{ijgmmp,vrflg,vncg,vsgg} (see also examples of 3, 4 and 5 dimensional
locally anisotropic black hole/ ellipsoid solutions Refs. \cite%
{vs5dbh,vbe1,vbe2}). Nevertheless, the above--mentioned types of locally
anisotropic solutions are not exactly tailored for the pseudo--Finsler
spacetimes but for more general nonholonomic configurations. Until the
present, there were not published any works on exact solutions for black
hole metrics and connections in Finsler gravity models.

A surprising result is that for certain classes of nonholonomic
distributions/ frames we can model (pseudo) Lagrange and Finsler like
geometries on (pseudo) Riemannian manifolds. In such cases, the metric and
connection structures can be constrained to solve usual gravitational field
equations in general relativity \cite{ijgmmp,vrflg,vbe1,vbe2}, or their
generalizations \cite{vncg,vsgg,vs5dbh}. So, the task to construct Finsler
black hole solutions is not only a formal one related to non--Riemannian
spacetimes but also presents a substantial interest in modelling locally
anisotropic black hole configurations, with generic off--diagonal metrics,
in Einstein gravity. Here we emphasize that locally anisotropic/
nonholonomic spacetimes (Finsler like and more general ones) defined as
exact solutions of standard Einstein equations are not subjected to
experimental constraints as in the case of modified gravity theories
constructed on (co) tangent bundles \cite{will,laem}.

It is complicated to find new classes of black hole solutions both in
Einstein gravity and modified gravity theories. Any type of such solutions
presents a substantial interest for possible applications in modern
astrophysics and cosmology. The subject of pseudo--Finsler black holes is
also interesting for various studies in mathematical physics as it is
connected to new methods of constructing exact solutions defining spacetimes
with generalized symmetries and nonlinear gravitational interactions \cite%
{ijgmmp,vrflg,vsgg}.

Recently, a series of works on Finsler--like analogous of gravity and
applications \cite{perl,mignemi,gibbons,sindoni,skak} were published
following various purposes in modern cosmology \cite{lin,chang,kstav},
string gravity \cite{vst1,vst2,mavr} and quantum gravity \cite{gls,vpla},
and alternative gravity theories and physical applications of nonholonomic
Ricci flows \cite{vnhrf}; see also monographs \cite%
{cart,rund,mats,as1,ma,bejf,vsp1,vsp2,bcs}, and references therein, on
\textquotedblright early\textquotedblright\ physical models with Finsler
geometries. Here we note also gravity and matter field interactions theories
with momentum type variables and higher order (co) tangent bundles: For
instance, such constructions were related to definition of nonholonomic
spinors, field equations and conservation laws on such spaces \cite%
{vsp1,vsp2} and various models with velocity and momentum/phase type
variables \cite{amelino,maueijo,ellis,mavr2,castro1,castro2}. To analyse
gamma ray bust delay times and relations to the geometry of momentum space
is proposed in \cite{friedel} but such a theory is not complete from a
geometric point of view because the approach does not include the nonlinear
connection structure and possible symplectomorphisms, see \cite{av}. It is
an important task to investigate if certain Finsler like gravitational
models (commutative and noncommutative ones, nonsymmetric metric
generalizations etc) may have, or not, black hole type solutions and to
understand when some (pseudo) Finsler metrics can be related to modern/
standard theories of gravity.

Exact solutions in gravity theories, including black hole metrics, carry a
great deal of information on gravitational theories. They can be considered
both for theoretical and, in many cases, experimental tests of physical
models. In our approach, we construct new classes of black hole/ ellipsoid
solutions generalizing similar ones in Einstein gravity; namely these are
the Schwarzschild analogs in pseudo--Finsler spacetimes, and analyze
possible physical implications.

In this article, we present and discuss the main properties of Finsler black
hole solutions generated following the so--called anholonomic frame method,
elaborated and developed in our previous works \cite%
{ijgmmp,vrflg,vncg,vsgg,vs5dbh,vs5dbh1,vbe1,vbe2}. We shall work explicitly
with Finsler spacetime configurations, and their nonholonomic transforms,
but not with generalized locally anisotropic and/or Lagrange--Finsler
structures in the bulk of our former constructions. Readers are recommended
to study preliminary the reviews \cite{ijgmmp,vrflg} on applications in
physics of the geometry of nonlinear connections and associated nonholonomic
frames and modelling of Lagrange--Finsler geometries on (pseudo) Riemannian
spacetimes.

We emphasize that in this work the Finsler geometry models are elaborated
for the canonical distinguished connection following geometric methods
developed in Refs. \cite{ma,bejf,vrflg,vsgg}. In our approach, we usually do
not work with the Chern connection \cite{bcs} (sometimes, called also Rund's
connection \cite{rund}) for Finsler spaces because this connection is metric
noncompatible, which results in a more sophisticate mathematical formalism
and have less physical motivations from the viewpoint of standard gravity
theories, see discussions in Ref. \cite{vrflg} and Introduction to \cite%
{vsgg}. For simplicity, this paper is oriented to keep the geometric and
physical constructions more closed to those in Einstein gravity, in Finsler
like variables, and four dimensional pseudo--Finsler analogs on nonholonomic
tangent bundles.

The paper is organized as follows:\ In section 2, we summarize some basic
concepts and formulas on modeling Finsler geometries on tangent bundles and
nonholonomic (pseudo) Riemannian manifolds. We consider the gravitational
field equations for Einstein--Finsler and nonholonomic Einstein spaces. In
section 3, we choose the metric ansatz and analyze four classes of exact
solutions parametrized by generic off--diagonal metrics with one Killing
vector symmetry defining stationary Einstein--Finsler configurations.

Solutions for pseudo--Finsler generalizations of the Schwarzschild metric
and nonholonomic ellipsoidal deformations of Einstein metrics are provided
in section 4. We discuss there how such metrics can be constructed on
tangent bundles/ Einstein manifolds. There are provided examples when black
hole solutions can be imbedded and/or nonholonomically mapped on
pseudo--Finsler spaces and/or deformed nonholonomically into exact solutions
of Einstein equations in general relativity. In section 5, we discuss and
conclude on obtained results.

Finally, we note that we cite only a series of Finsler works which may have
implications for standard theories of gravity and high energy physics, in
the spirit of reviews \cite{vrflg,ijgmmp} and monograph \cite{vsgg}. In a
recent paper \cite{vgsol}, we proved that the Einstein equations, both with
the canonical d--connection and Levi--Civita connection, can be solved in
very general form for arbitrary dimensions. All solutions considered in this
work correspond to explicit classes of generating and integration functions
generating ellipsoid and/or solitonic configurations. They provide some
possible applications (for certain types of nonholonomic distributions
defining new classes of black hole objects) in Finsler and Einstein gravity.
More details on the geometry of Finsler spaces, generalizations and
"nonstandard" applications in physics are presented in \cite%
{cart,rund,mats,ma,bejf,vsp1,vsp2,bcs} and references therein.

\section{Einstein--Finsler Manifolds}

Let us consider a four dimensional (4-d) manifold $\mathbf{V}$ of necessary
smooth class (in brief, we shall use the terms space/ or spacetime, for
corresponding positive/ negative signatures of metrics). We shall model
Finsler geometries with geometric/physical objects defined on $\mathbf{V}$
as exact solutions of certain gravitational field equations in general
relativity or modified for Finsler metrics and connections. Such spacetimes
are enabled with a conventional $2+2$ splitting (defined by a nonholonomic,
equivalently, anholonomic/nonintegrable, distribution), when local
coordinates $u=(x,y)$ on an open region $U\subset \mathbf{V}$ are labelled
in the form $u^{\alpha }=(x^{i},y^{a}),$ with indices of type $i,j,k,...=1,2$
and $a,b,c...=3,4.$ For tensor like objects on $\mathbf{V},$ their
coefficients will be considered with respect to a general (non--coordinate)
local basis $e_{\alpha }=(e_{i},e_{a}).$\footnote{%
Primed (double primed, underlined etc) indices, for instance $\alpha
^{\prime }=(i^{\prime },a^{\prime }),$ $\beta ^{\prime \prime }=(j^{\prime
\prime },b^{\prime \prime }),\underline{\gamma }=(\underline{k},\underline{c}%
),...$ will be used for labelling coordinates with respect to a different
local basis $e_{\alpha ^{\prime }}=(e_{i^{\prime }},e_{a^{\prime }}),$ or
its dual $e^{\alpha ^{\prime }}=(e^{i^{\prime }},e^{a^{\prime }}).$}

There are two different ways to model Finsler geometries as solutions of
gravitational field equations in general relativity and/or modified
theories:\ The first, original (standard) one \cite{cart,rund,mats,ma,bcs},
is to consider $\mathbf{V}=TM$ \ for the total space of a tangent bundle $%
\left( TM,\pi ,M\right) $ on a two dimensional (2--d) base manifold $M.$ In
this approach, the values $x^{i}$ are local coordinates on $M$ (a
low--dimensional space/ spacetime) and $y^{a}$ are fiber/velocity type
coordinates. In such a case, the geometric constructions and physical models
are performed on tangent bundles which results in generalizations/
violations of the local Lorentz invariance, non--Riemannian locally
anisotropic gravitational effects etc. Positively, Finsler gravity models
constructed on $TM$ should be considered as modifications of the Einstein
theory.

Alternatively, we can consider that $\mathbf{V}=V$ is a 4--d nonholonomic
manifold (in particular, a pseudo--Riemannian one) with local fibered
structure\footnote{%
A pair $(\mathbf{V},\mathcal{N})$, where $\mathbf{V}$ is a manifold and $%
\mathcal{N}$ is a nonintegrable distribution on $\mathbf{V}$, is called a
nonholonomic manifold. In this work, boldface symbols will be used for
nonholonomic manifolds/ bundles and geometric objects on such spaces.}, as
we discuss in \cite{vrflg,ijgmmp,vsgg}. For such constructions, we treat $%
x^{i}$ a nd $y^{a},$ respectively, as conventional horizontal/ nonholonomic
(h) and vertical / holonomic (v) coordinates (both types of such coordinates
can be time-- or space--like ones). On nonholonomic (pseudo) Riemannian
manifolds, we can introduce Finsler and/or almost K\" ahler like
variables/cordinates \cite{vrflg,vpla,vbrane}.\footnote{\label{fnoteprc} We
emphasize that the spacetime signature may be encoded formally into certain
systems of frame (vielbein) coefficients and coordinates, some of them being
proportional to the imaginary unity $i,$ when $i^{2}=-1.$ For instance, on a
local tangent Minkowski space of signature $(-,+,+,+),$ we can chose $%
e_{1^{\prime }}=i\partial /\partial u^{1^{\prime }},$ where $i$ is the
imaginary unity, $i^{2}=-1,$ and write $e_{\alpha ^{\prime }}=(i\partial
/\partial u^{1^{\prime }},\partial /\partial u^{2^{\prime }},\partial
/\partial u^{3^{\prime }},\partial /\partial u^{4^{\prime }}).$ Such formal
local Euclidean coordinates were used in the past in a number of textbooks
on relativity theory (see, for instance, \cite{landau,moller}). This is
useful for some purposes of analogous modelling of gravity theories as
effective Lagrange mechanics, or Finsler like, geometries, but this does not
mean that we work with a complexification of classical spacetimes, see
discussion in Ref. \cite{vrflg}. The term ''pseudo--Finsler'' was also
considered more recently for some analogous gravity like models (see, for
instance, \cite{skak}).} In a more general context, we can model by
nonholonomic distributions / frames on (pseudo) Riemannian/ Einstein
manifolds various types of (pseudo) Finsler geometries when coordinates of
type $y^a$ are not "velocities" but certain nonholonomically constrained
variables, for instance, in general relativity.

\subsection{(Pseudo) Finsler/ Riemannian metrics and Finsler variables}

\label{ssfg}A standard two dimensional Finsler space $\ ^{2}F(M,F(x,y))$ is
defined on a tangent bundle $TM,$ where $M,\dim M=2,$ is the base manifold
being differentiable of class $C^{\infty }.$ One considers: 1) a fundamental
real Finsler (generating) function $F(u)=F(x,y)=F(x^{i},y^{a})>0$ if $y\neq
0 $ and homogeneous of type $F(x,\lambda y)=|\lambda |F(x,y),$ for any
nonzero $\lambda \in \mathbb{R},$ with positively definite Hessian
\begin{equation}
\ f_{ab}=\frac{1}{2}\frac{\partial ^{2}F^{2}}{\partial y^{a}\partial y^{b}},
\label{efm}
\end{equation}%
when $\det |\ f_{ab}|\neq 0.$ In order to construct completely a geometric
model (Finsler geometry) on $TM,$ we have to chose 2) a nonlinear connection
(N--connection) structure $\mathbf{N}$ on $TM$ defined by a nonholonomic
distribution (Whitney sum)%
\begin{equation}
TTM=hTM\oplus vTM  \label{whitney}
\end{equation}%
with splitting into conventional horizontal (h), $hTM,$ and vertical (v), $%
vTM,$ subspaces and 3) a linear connection structure which is convenient to
be defined in N--adapted form, i.e. preserving the splitting (\ref{whitney}%
), called distinguished connection (in brief, d--connection) and denoted $%
\mathbf{D}=(hD,vD),$ or $\mathbf{D}_{\alpha }=(D_{i},D_{a}).$

For a Finsler space, it is possible to construct the canonical (Cartan)
N--connection $\ ^{c}\mathbf{N}=\{\ ^{c}N_{i}^{a}\}$ completely defined by
an effective Lagrangian $L=F^{2}$ in such a form that the corresponding
semi--spray configuration is described by nonlinear geodesic equations. In
their turn, the geodesic equations are equivalent to the Euler--Lagrange
equations for $L$ (see details, for instance, in Refs. \cite{ma,vrflg,vsgg};
for ''pseudo'' configurations, this mechanical analogy is a formal one, with
some ''imaginary'' coordinates). One introduces
\begin{equation}
\ ^{c}N_{i}^{a}=\frac{\partial G^{a}}{\partial y^{2+i}},  \label{clnc}
\end{equation}%
for $\ G^{a}=\frac{1}{4}\ f^{a\ 2+i}\left( \frac{\partial ^{2}L}{\partial
y^{2+i}\partial x^{k}}y^{2+k}-\frac{\partial L}{\partial x^{i}}\right) ,$
where $\ f^{ab}$ is inverse to $\ f_{ab}$ (\ref{efm}) and respective
contractions of $h$-- and $v$--indices, $\ i,j,...$ and $a,b...,$ are
performed following the rule: we can write, for instance, an up $v$--index $%
a $ as $a=2+i$ and contract it with a low index $i=1,2.$ Briefly, for spaces
of even dimension, we can write $y^{i}$ instead of $y^{2+i},$ or $y^{a}.$

The values (\ref{efm}) and (\ref{clnc}) allow us to define the canonical
Sasaki type metric (d--metric, equivalently, metric d--tensor)
\begin{equation}
\ \mathbf{f}=\ f_{ij}dx^{i}\otimes dx^{j}+\ f_{ab}\ ^{c}\mathbf{e}%
^{a}\otimes \ ^{c}\mathbf{e}^{b},\ \ ^{c}\mathbf{e}^{a}=dy^{a}+\
^{c}N_{i}^{a}dx^{i},  \label{fsm}
\end{equation}%
where $f_{ij}=\ f_{2+i\ 2+j}.$ For (pseudo) Finsler spaces, we have to
consider that Hessian (\ref{efm}) is not positively definite and that (\ref%
{fsm}) has locally a (pseudo) Euclidean signature.\footnote{%
It is possible to construct Finsler models on total bundle of $TM$ for
various types of quadratic forms which, for instance, are completely
homogeneous on $y$--variables\cite{bm,bcs}. This results in more
sophisticate geometric constructions with less obvious physical
applications. In our works, we use the Sasaki form as the most ''simple''
example both in geometry in physics.}

Instead of a tangent bundle $TM,$ we can model a Finsler geometry on a 4--d
(pseudo) Riemannian manifold $\ ^{4}V$ (of necessary smooth class) if we
consider that such a manifold is enabled with a nonholonomic distribution of
type (\ref{whitney}), where $TM$ is substituted by $\ ^{4}V.$ The
coefficients of a general (pseudo) Riemannian metric $g=g_{\alpha ^{\prime
}\beta ^{\prime }}e^{\alpha ^{\prime }}\otimes e^{\beta ^{\prime }}$ on $\
^{4}V,$ \ for $e^{\alpha ^{\prime }}=(e^{i^{\prime }},e^{a^{\prime }})=e_{\
\alpha }^{\alpha ^{\prime }}(u)du^{\alpha },$ can be parametrized in a form
adapted to a nonholonomic $2+2$ splitting induced by a (formally introduced)
Finsler generating function. In such cases, we have \
\begin{equation}
g=\mathbf{g}=g_{i^{\prime }j^{\prime }}(u)e^{i^{\prime }}\otimes
e^{j^{\prime }}+h_{a^{\prime }b^{\prime }}(u)\mathbf{e}^{a^{\prime }}\otimes
\mathbf{e}^{b^{\prime }},\ e^{a^{\prime }}=\mathbf{e}^{a^{\prime
}}+N_{i^{\prime }}^{a^{\prime }}(u)e^{i^{\prime }},  \label{gpsm}
\end{equation}%
when the values $\mathbf{g}_{\alpha ^{\prime }\beta ^{\prime
}}=[g_{i^{\prime }j^{\prime }},h_{a^{\prime }b^{\prime }}]$ are related by
transforms
\begin{equation}
\mathbf{g}_{\alpha ^{\prime }\beta ^{\prime }}e_{\ \alpha }^{\alpha ^{\prime
}}e_{\ \beta }^{\beta ^{\prime }}=\mathbf{f}_{\alpha \beta }  \label{algeq}
\end{equation}%
to a (pseudo--Finsler) metric $\ \mathbf{f}_{\alpha \beta }=\ [\ f_{ij},\
f_{ab}]$ (\ref{fsm}) and corresponding N--adapted dual canonical basis $\
^{c}e^{\alpha }=(dx^{i},\ ^{c}e^{a}).$\footnote{%
For any given values $\mathbf{g}_{\alpha ^{\prime }\beta ^{\prime }}$ and $~%
\mathbf{f}_{\alpha \beta },$ we have to solve a system of quadratic
algebraic equations (\ref{algeq}) with unknown variables $e_{\ \alpha
}^{\alpha ^{\prime }}.$ How to define in explicit form such frame \
coefficients (vierbeins) and coordinates we discuss in detail, for instance,
in Ref. \cite{ijgmmp}.} Usually, for certain formally diagonalized
representations of h- and v--components of metrics, a subset of such
coefficients can be taken to be zero, for given values $[g_{i^{\prime
}j^{\prime }},h_{a^{\prime }b^{\prime }},N_{i^{\prime }}^{a^{\prime }}]$ and
$\ [\ f_{ij},\ f_{ab},\ \ ^{c}N_{i}^{a}],$ when 
$N_{i^{\prime }}^{a^{\prime }}=e_{i^{\prime }}^{\ i}e_{\ a}^{a^{\prime }}\ \
^{c}N_{i}^{a},$ 
for $e_{i^{\prime }}^{\ i}$ being inverse to $e_{\ i}^{i^{\prime }}.$

In our unified geometric approach (both for tangent bundles and nonholonomic
manifolds), we work on a general spacetime $\mathbf{V}$ enabled with a
N--connecti\-on structure $\mathbf{N}$ defined by a Whitney sum of type (\ref%
{whitney}), with local coefficients $N_{i}^{a}(u),$ when $\mathbf{N}%
=N_{i}^{a}(u)dx^{i}\otimes \frac{\partial }{\partial y^{a}}.$ Such a
nonholonomic distribution (called also N--anholonomic, see \cite{vrflg,vsgg}%
; it is also used the term N--anholonomic manifold) states a frame
(vielbein) structure which is linear on N--connection coefficients,
\begin{eqnarray}
\mathbf{e}_{\nu } &=&\left( \mathbf{e}_{i}=\frac{\partial }{\partial x^{i}}%
-N_{i}^{a}(u)\frac{\partial }{\partial y^{a}},e_{a}=\frac{\partial }{%
\partial y^{a}}\right) ,  \notag \\
\mathbf{e}^{\mu } &=&\left( e^{i}=dx^{i},\mathbf{e}%
^{a}=dy^{a}+N_{i}^{a}(u)dx^{i}\right) .  \label{ddif}
\end{eqnarray}%
The vielbeins (\ref{ddif}) satisfy the nonholonomy relations
\begin{equation}
\lbrack \mathbf{e}_{\alpha },\mathbf{e}_{\beta }]=\mathbf{e}_{\alpha }%
\mathbf{e}_{\beta }-\mathbf{e}_{\beta }\mathbf{e}_{\alpha }=w_{\alpha \beta
}^{\gamma }\mathbf{e}_{\gamma }  \label{anhrel}
\end{equation}%
with (antisymmetric) nontrivial anholonomy coefficients $w_{ia}^{b}=\partial
_{a}N_{i}^{b}$ and $w_{ji}^{a}=\Omega _{ij}^{a},$ where $\Omega _{ij}^{a}=%
\mathbf{e}_{j}\left( N_{i}^{a}\right) -\mathbf{e}_{i}\left( N_{j}^{a}\right)
$ \ are the coefficients of N--connection curvature. The particular
holonomic/ integrable case is selected by the integrability conditions $%
w_{\alpha \beta }^{\gamma }=0.$\footnote{%
We use boldface symbols for spaces (and geometric objects on such spaces)
enabled with N--connection structure.}

For geometric constructions on $\mathbf{V}$ and physical applications, it is
convenient to work with a class of linear connections $\mathbf{D}$ which
under parallelism preserve a prescribed distribution (\ref{whitney}). Such
connections are called N--adapted (or, equivalently, distinguished
connections; in brief, d--connections). The simplest way to perform
computations with d--connections is to use N--adapted differential 1--forms
of type
\begin{equation}
\mathbf{\Gamma }_{\ \beta }^{\alpha }=\mathbf{\Gamma }_{\ \beta \gamma
}^{\alpha }\mathbf{e}^{\gamma }.  \label{dconf}
\end{equation}%
For instance, the torsion of a d--connection $\mathbf{D}$ is computed in
N--adapted form as 
$\mathcal{T}^{\alpha }\doteqdot \mathbf{De}^{\alpha }=d\mathbf{e}^{\alpha }+%
\mathbf{\Gamma }_{\ \beta }^{\alpha }\wedge \mathbf{e}^{\beta },$
which result in coefficients $\mathbf{T}_{\ \beta \gamma }^{\alpha }=\{T_{\
jk}^{i},T_{\ ja}^{i},T_{\ ji}^{a},T_{\ bi}^{a},T_{\ bc}^{a}\},$ where
\begin{eqnarray}
T_{\ jk}^{i} &=&L_{\ jk}^{i}-L_{\ kj}^{i},\ T_{\ ja}^{i}=-T_{\ aj}^{i}=C_{\
ja}^{i},\ T_{\ ji}^{a}=\Omega _{\ ji}^{a},\   \notag \\
T_{\ bi}^{a} &=&-T_{\ ib}^{a}=\frac{\partial N_{i}^{a}}{\partial y^{b}}-L_{\
bi}^{a},\ T_{\ bc}^{a}=C_{\ bc}^{a}-C_{\ cb}^{a}.  \label{dtors}
\end{eqnarray}

For any $\mathbf{g}$ on a N--anholonomic manifold $\mathbf{V,}$ there is a
unique metric compatible canonical d--connection $\widehat{\mathbf{D}}$ when
$\widehat{\mathbf{D}}\mathbf{g=}0$ and ''pure'' horizontal and vertical
torsion coefficients are zero, i. e. $\widehat{T}_{\ jk}^{i}=0$ and $%
\widehat{T}_{\ bc}^{a}=0,$ see formulas (\ref{dtors}). Locally, a canonical
d--connection $\widehat{\mathbf{D}}$ is determined by its coefficients $%
\widehat{\mathbf{\Gamma }}_{\ \alpha \beta }^{\gamma }=\left( \widehat{L}%
_{jk}^{i},\widehat{L}_{bk}^{a},\widehat{C}_{jc}^{i},\widehat{C}%
_{bc}^{a}\right) ,$ where
\begin{eqnarray}
\widehat{L}_{jk}^{i} &=&\frac{1}{2}g^{ir}\left(
e_{k}g_{jr}+e_{j}g_{kr}-e_{r}g_{jk}\right) ,  \label{candcon} \\
\widehat{L}_{bk}^{a} &=&e_{b}(N_{k}^{a})+\frac{1}{2}h^{ac}\left(
e_{k}h_{bc}-h_{dc}\ e_{b}N_{k}^{d}-h_{db}\ e_{c}N_{k}^{d}\right) ,  \notag \\
\widehat{C}_{jc}^{i} &=&\frac{1}{2}g^{ik}e_{c}g_{jk},\ \widehat{C}_{bc}^{a}=%
\frac{1}{2}h^{ad}\left( e_{c}h_{bd}+e_{c}h_{cd}-e_{d}h_{bc}\right) .  \notag
\end{eqnarray}%
We emphasize that, in general, $\widehat{T}_{\ ja}^{i},\widehat{T}_{\
ji}^{a} $ and $\widehat{T}_{\ bi}^{a}$ are not zero, but such nontrivial
components of torsion are induced by some coefficients of a general
off--diagonal metric $\mathbf{g}_{\alpha \beta },$ see explicit formulas in
Refs. \cite{vrflg,vsgg}. Distinguished connections were originally
introduced for Finsler and Lagrange models on tangent bundles (see, for
instance, \cite{ma,bcs,mats,as1,cart}). They can be also defined on
(pseudo)\ Riemannian manifolds by adapting the geometric constructions with
respect to N--anholonomic structures. For instance, a $\widehat{\mathbf{D}}$
on $\mathbf{V}$ is completely defined by the metric structure.

Usually, in (pseudo) Riemannian geometry, it is considered another
''standard'' connection, the so--called Levi--Civita linear connection, $\
\nabla =\{\Gamma _{\beta \gamma }^{\alpha }\}$ which is uniquely defined by
a metric structure $\mathbf{g}$ following the conditions $\ ^{\nabla }%
\mathcal{T}=0$ and $\nabla \mathbf{g}=0.$ Any geometric construction for the
canonical d--connection $\widehat{\mathbf{D}}$ can be re--defined
equivalently into a similar one with the Levi--Civita connection $\nabla $
following formula
\begin{equation}
\Gamma _{\ \alpha \beta }^{\gamma }=\widehat{\mathbf{\Gamma }}_{\ \alpha
\beta }^{\gamma }+Z_{\ \alpha \beta }^{\gamma },  \label{deflc}
\end{equation}%
where the distortion tensor $\ Z_{\ \alpha \beta }^{\gamma }$ is constructed
in a unique form from the coefficients of a metric $\mathbf{g}_{\alpha \beta
},$%
{\small 
\begin{eqnarray}
Z_{jk}^{a} &=&-\widehat{C}_{jb}^{i}g_{ik}h^{ab}-\frac{1}{2}\Omega
_{jk}^{a},~Z_{bk}^{i}=\frac{1}{2}\Omega _{jk}^{c}h_{cb}g^{ji}-\Xi _{jk}^{ih}~%
\widehat{C}_{hb}^{j},  \notag \\
Z_{bk}^{a} &=&~^{+}\Xi _{cd}^{ab}\widehat{T}_{bk}^{c},\ Z_{kb}^{i}=\frac{1}{2%
}\Omega _{jk}^{a}h_{cb}g^{ji}+\Xi _{jk}^{ih}~\widehat{C}_{hb}^{j},\
Z_{jk}^{i}=0,  \label{deft} \\
\ Z_{jb}^{a} &=&-~^{-}\Xi _{cb}^{ad}\widehat{T}_{dj}^{c},\ Z_{bc}^{a}=0,\
Z_{ab}^{i}=-\frac{g^{ij}}{2}\left[ \widehat{T}_{aj}^{c}h_{cb}+\widehat{T}%
_{bj}^{c}h_{ca}\right] ,  \notag
\end{eqnarray}%
}
for $\ \Xi _{jk}^{ih}=\frac{1}{2}(\delta _{j}^{i}\delta
_{k}^{h}-g_{jk}g^{ih}),~^{\pm }\Xi _{cd}^{ab}=\frac{1}{2}(\delta
_{c}^{a}\delta _{d}^{b}+h_{cd}h^{ab})$ and$~\widehat{T}_{\ ja}^{c}=\widehat{L%
}_{aj}^{c}-e_{a}(N_{j}^{c}).$

\vskip5pt

A (pseudo) Riemannian geometry is completely defined by only one fundamental
geometric object which is the metric structure $\mathbf{g.}$ It determines a
unique metric compatible and torsionless Levi--Civita connection $\nabla .$
In order to construct a model of (pseudo) Finsler geometry, we need a
generating fundamental Finsler function $F(x,y)$ which in certain canonical
approaches generates three fundamental geometric objects: 1) a N--connection
$\mathbf{N},$ 2) a Sasaki type metric $\mathbf{f}_{\alpha \beta }$ and 3) a
d--connection $\mathbf{D.}$ Using distortions of d--connections, $\ \nabla =%
\widehat{\mathbf{D}}+\mathbf{Z}$ (\ref{deflc}), and nonhlonomic frame
transform (\ref{algeq}), we can encode a canonical (pseudo) Finsler geometry
$(F,\ \mathbf{f,}\ \ ^{c}\mathbf{N,}\widehat{\mathbf{D}})$ \ into a (pseudo)
Riemannian configuration $(\mathbf{g,}\nabla ),$ with $\mathbf{g}$ induced
by a Finsler metric $\ \mathbf{f}$ (\ref{fsm}). Inversely, for any (pseudo)
Riemannian space, we can chose a nonholonmic 2+2 distribution induced by a
necessary type of Finsler generating function $F(x,y)$ and express any data $%
(\mathbf{g,}\nabla )$ into, for instance, canonical (pseudo) Finsler ones.
In this way we introduce (nonholonomic) Finsler variables on a (pseudo)
Riemannian manifold and deform nonholonomically the linear connection
structure, $\ \nabla \rightarrow \widehat{\mathbf{D}}$ in order to model a
Finsler geometry by a corresponding nonholonomic distribution. For such
geometric constructions, we can work equivalently with two types of linear
connections, $\nabla $ and/or $\widehat{\mathbf{D}},$ because all values $%
\nabla ,\widehat{\mathbf{D}}$ and$\ \mathbf{Z}$ in (\ref{deflc}), are
determined by the same metric structure $\mathbf{g=}\ \mathbf{f.}$

The main conclusion of this section is that geometrically both types of
(pseudo) Riemannian and (pseudo) Finsler spaces with metric compatible
linear connections completely defined by a prescribed metric structure can
be modelled equivalently by nonholonomic distributions/ deformations. For
instance, we can consider that a (pseudo) Finsler geometry is modelled on a
tangent bundle, when v--coordinates $y^{a}$ are of ''velocity'' type. This
is a very different (from physical point of view, but with a formal
equivalence of geometric formulas) from Finsler spacetime models on
nonholonomic (pseudo) Riemannian manifolds, when v--coordinates $y^{a}$ are
certain space/time ones subjected to nonholonomic constraints.

\subsection{Einstein equations on (pseudo) Finsler/ nonholo\-no\-mic
spacetimes}

For any d--connection $\mathbf{D},$ we can compute the nontrivial N--adapted
components of curvature, $\mathbf{\mathbf{R}_{\ \beta \gamma \delta
}^{\alpha }=}\{R_{\ hjk}^{i},R_{\ bjk}^{a},R_{\ jka}^{i},R_{\ bka}^{c},R_{\
jbc}^{i},R_{\ bcd}^{a}\},$\footnote{%
in explicit form, the formula for such coefficients can be found in Refs.
\cite{ma,vrflg,vsgg}} following a formal differential form calculus,
\begin{equation*}
\mathcal{R}_{~\beta }^{\alpha }\doteqdot \mathbf{D\Gamma }_{\ \beta
}^{\alpha }=d\mathbf{\Gamma }_{\ \beta }^{\alpha }-\mathbf{\Gamma }_{\ \beta
}^{\gamma }\wedge \mathbf{\Gamma }_{\ \gamma }^{\alpha }=\mathbf{R}_{\ \beta
\gamma \delta }^{\alpha }\mathbf{e}^{\gamma }\wedge \mathbf{e}^{\delta },
\end{equation*}%
Contracting respectively the components of curvature, one proves that the
Ricci tensor $\mathbf{R}_{\alpha \beta }\doteqdot \mathbf{R}_{\ \alpha \beta
\tau }^{\tau }$ is characterized by h- v--components, $\mathbf{R}_{\alpha
\beta }=\{R_{ij},R_{ia},\ R_{ai},\ R_{ab}\},$ for $R_{ij}\doteqdot R_{\
ijk}^{k},\ \ R_{ia}\doteqdot -R_{\ ika}^{k},\ R_{ai}\doteqdot R_{\
aib}^{b},\ R_{ab}\doteqdot R_{\ abc}^{c}.$ The scalar curvature of a
d--connection $\mathbf{D}$ is standardly defined $\ \ ^{s}\mathbf{R}%
\doteqdot \mathbf{g}^{\alpha \beta }\mathbf{R}_{\alpha \beta
}=g^{ij}R_{ij}+h^{ab}R_{ab}.$

The Einstein tensor of d--connection is by definition
\begin{equation}
\mathbf{E}_{\alpha \beta }=\mathbf{R}_{\alpha \beta }-\frac{1}{2}\mathbf{g}%
_{\alpha \beta }\ ^{s}\mathbf{R}.  \label{enstdt}
\end{equation}%
It defines a nonholonomic Einstein configuration which is alternative to the
standard one constructed for the Levi--Civita connection.\footnote{%
The Levi--Civita connection and related Christoffel symbols are considered
as the standard geometric objects used in general relativity. It should be
noted here that the Einstein theory can be formulated equivalently in terms
of any linear connection if such a connection is completely determined by
the (pseudo) Riemannian metric structure. Using distortions of linear
connections of type (\ref{deflc}), we can re--express all fundamental
physical values and equations in general relativity in terms of certain
d--objects and, inversely, in terms of standard (pseudo) Riemannian,
Levi--Civita, tensor calculus.}

Working with Einstein tensors for different linear connections uniquely
defined by the same metric structure, we construct (in general) different/
alternative models of gravity. The corresponding gravitational field
equations are with different classes of exact solutions. Nevertheless, it is
possible to impose certain types of well defined constraints when the
solutions from one theory will be transformed into solutions from another
type theory. For instance, having constructed a solution of Einstein
equations for $\widehat{\mathbf{D}},$ we can restrict it to be a solution
for $\nabla $ if the coefficients of metric to satisfy the condition $Z_{\
\alpha \beta }^{\gamma }=0,$ see (\ref{deft}).

\subsubsection{Einstein equations in Finsler variables}

In general relativity, the Einstein equations are postulated in the form
\begin{equation}
\ E_{\alpha \beta }=\varkappa \ \Upsilon _{\alpha \beta },  \label{eeqlc}
\end{equation}%
where $\ E_{\alpha \beta }$ is the Einstein tensor for the Levi--Civita
connection $\nabla ,\varkappa $ is the gravitational constant and the
energy--momentum tensor $\ \Upsilon _{\alpha \beta }$ is constructed for
matter fields using $(\mathbf{g,}\nabla ).$ Having prescribed a generating
function $\mathbf{f},$ these equations can be written equivalently in
Finsler variables, when $(\mathbf{g,}\nabla )\rightarrow (F:\ \mathbf{f,}\
^{c}\mathbf{N,}\widehat{\mathbf{D}}),$ following the distortion of
connections (\ref{deflc}). Using the Einstein tensor (\ref{enstdt}) computed
for $\widehat{\mathbf{D}},$ we represent the gravitational field equations
in general relativity in the form
\begin{equation}
\ \widehat{\mathbf{E}}_{\alpha \beta }=\varkappa \ ^{c}\widehat{\mathbf{%
\Upsilon }}_{\alpha \beta },  \label{eeqlcf}
\end{equation}%
with a new (distorted) source $\ ^{c}\widehat{\mathbf{\Upsilon }}_{\alpha
\beta }=\Upsilon _{\alpha \beta }-\widehat{\mathbf{Z}}_{\ \alpha \beta }$
determined by $\ \Upsilon _{\alpha \beta }$ from (\ref{eeqlc}), rewritten in
variables $(\mathbf{g,}\ \widehat{\mathbf{D}}),$ and by the canonical
distortion of the Ricci tensor $\ \widehat{\mathbf{Z}}_{\ \alpha \beta }.$
This tensor is computed as $\ \widehat{\mathbf{Z}}_{\ \beta \gamma
}\doteqdot \ \widehat{\mathbf{Z}}_{\ \beta \gamma \alpha }^{\alpha }$ for $%
\widehat{\mathbf{R}}_{\ \beta \gamma \delta }^{\alpha }=\ R_{\ \beta \gamma
\delta }^{\alpha }+\ \widehat{\mathbf{Z}}_{\ \beta \gamma \delta }^{\alpha }$
induced by deformations (\ref{deflc}). So, the source $\ ^{c}\widehat{%
\mathbf{\Upsilon }}_{\alpha \beta }$ is such way constructed that (\ref%
{eeqlcf}) is equivalent to (\ref{eeqlc}) even such gravitational field
equations are written for different connections determined by the same
metric structure.

\subsubsection{A  Finsler gravity model with Einstein tensor for $%
\widehat{\mathbf{D}}$}

On a (pseudo) Finsler spacetime, the geometric constructions should be
adapted to a N--connection splitting. To satisfy this condition we should
not work with the Levi--Civita connection $\nabla $ but, for instance, with $%
\widehat{\mathbf{D}},$ or another type of d--connection $\mathbf{D}$. In a
''canonical'' d--connection approach, we are ''obliged'' to work with the
Einstein tensor $\ \widehat{\mathbf{E}}_{\alpha \beta }.$\footnote{%
On a Finsler spacetime, with respect to a N--adapted frame and fixed system
of coordinates, we can always compute $\nabla $ from $\widehat{\mathbf{D}}.$
If we work in a formal way with $\nabla,$ we ''hide'' the N--adapted
structure which is fundamental in Finsler gravity. To introduce Finsler
variables in general relativity is an ''option'' which may help, for
instance, to solve some systems of equations, define some symmetries of
interactions, elaborate new schemes of quantization etc. On a ''true''
Finsler spacetime (on a tangent bundle), we are positively obliged to work
with $\widehat{\mathbf{D}}$ and N--adapted geometric objects/ constructions.}

Following certain analogy with general relativity (but elaborating
geometric/physical models on tangent bundles), we can postulate the field
equations
\begin{equation}
\ \widehat{\mathbf{E}}_{\alpha \beta }=\varkappa \ \widehat{\mathbf{\Upsilon
}}_{\alpha \beta }  \label{eeqcdc}
\end{equation}%
constructed for the Einstein d--tensor $\ \widehat{\mathbf{E}}_{\alpha \beta
}$ and any  $\ \widehat{\mathbf{\Upsilon }}_{\alpha \beta }.$ Such equations
can be used as fundamental field equations for Finsler gravity theories.
From a formal point of view, they do not have a rigorous physically
motivation like in general relativity\footnote{%
this is because we do not have any experimental data to support a principle
of equivalence for spacetime models on tangent bundles when metrics depend
on \textquotedblright velocities\textquotedblright }, but can be derived
geometrically following the same constructions for d--connections as for the
Levi--Civita connection.

We can elaborate a N--adapted variational approach for theories with $\nabla
\rightarrow \ \widehat{\mathbf{D}}$ for the same metric structure on $TM$
using N--elongated partial derivatives and co--frames  (\ref{ddif}). All
constructions can be equivalently re--defined with respect to coordinate
frames but  in such a case the N--connection structure will be "hidden" in
generic off--diagonal coefficients of metric. Variational "proofs" for the
Einstein equations for $\nabla ,$ i.e.  (\ref{eeqlc}) and the nonholonomic
canonical deformation to (\ref{eeqcdc}) are equivalent if both connections
are related by a distortion (\ref{deflc}), $\nabla =\widehat{\mathbf{D}}+%
\widehat{\mathbf{Z}}$ when all geometric objects $\nabla ,\widehat{\mathbf{D}%
}$ and $\widehat{\mathbf{Z}}$ are determined by the same metric structure $%
\mathbf{g}.$ The constructions can be generalized for Finsler like models of
gravity with arbitrary metric compatible of noncompatible d--connection $%
\mathbf{D}.$ Using a N--adapted Palatini method,  subclasses of generalized
Finsler like theories can be constructed for any data $(\mathbf{g,N,D}),$
see details in Ref.  \cite{vsgg}.

We note that a formal definition of $\ \widehat{\mathbf{\Upsilon }}_{\alpha
\beta }$ by using the same principles for $\widehat{\mathbf{D}},$ instead of
$\nabla ,$ like in Einstein gravity, is well defined geometrically and may
present certain interest. For instance, we can consider such an approach for
elaborating models with local anisotropies on velocities in modern cosmology
and/or continuous media mechanics.

\subsubsection{Nonholonomic Einstein and Einstein--Finsler spaces}

For our further purposes, we consider a particular case of equations (\ref%
{eeqcdc}), when the source $\ \widehat{\mathbf{\Upsilon }}_{\alpha \beta }$
is such way chosen that
\begin{equation}
\widehat{R}_{\ j}^{i}=\ ^{h}\lambda (u)\delta _{\ j}^{i},\ \widehat{R}_{\
b}^{a}=\ ^{v}\lambda (u)\delta _{\ b}^{a},\ \widehat{R}_{ia}=\ \widehat{R}%
_{ai}=0,  \label{eeqcdcc}
\end{equation}%
where $\ ^{h}\lambda (u)$ and $\ ^{v}\lambda (u)$ are some locally
anisotropic h-- and v--polarizations of the cosmological constant. If we
take $\ \widehat{\mathbf{D}}$ on a tangent bundle, $\mathbf{V}=TM,$ such
equations define a class of Einstein--Finsler spaces. Choosing $\mathbf{V}=\
^{4}V,$ we say that the solutions of equations (\ref{eeqcdcc}) are for
nonholonomic Einstein \ manifolds. We suggest to use such terms because for $%
\ \widehat{\mathbf{D}}\rightarrow \nabla $ and $\ ^{h}\lambda =$ $\
^{v}\lambda =\lambda =const$ we obtain the usual gravitational field
equations with cosmological constant,
\begin{equation}
\ R_{\alpha \beta }=\lambda g_{\alpha \beta },  \label{eeqlcc}
\end{equation}%
defining the concept of Einstein spaces/manifolds.

There are two important arguments to introduce and analyze in this work some
classes of exact solutions (which seem to have physical importance) of
equations (\ref{eeqcdcc}): The first one is that such equations allow us to
construct generalizations of black hole metrics for Finsler spacetimes
(depending on our choice, they can be similarly considered on tangent
bundles or on nonholonomic, Einstein manifolds). It may be physically
important to construct, for instance, Finsler analogs of the Schwarzschild
solutions.\footnote{%
As a Finsler black hole, we shall consider any black like solution of
(Finsler gravity) equations (\ref{eeqcdcc}). For well defined conditions,
such a solution may be restricted to solve the Einstein equations (\ref%
{eeqlcc})) but (in general) it contains certain generic off--diagonal terms
in metrics. All black hole solutions can be represented in Finsler variables
using transforms (\ref{algeq}) to a pseudo--Finsler metric with nontrivial
N--connection structure. If we choose $y$--variables for $TM,$ such Finsler
black holes depend on a "velocity" type coordinate. For nonholonomic
Einsteing spaces, we mimic a Finler like black hole configuration with a
nonholonomic $y$--variable.} In such cases, we have to solve modified
Einstein equations for $\widehat{\mathbf{D}}.$ The second argument is that
we can solve the equations (\ref{eeqcdcc}) analytically in very general form
for arbitrary dimensions, see recent results in Ref. \cite{vgsol} (we were
not able to generate such solutions working directly with the Levi--Civita
connection and equations (\ref{eeqlcc})). We can constrain such general
solutions and select some explicit nonholonomic configurations when the
distortion tensor vanishes,$\ Z_{\ \alpha \beta }^{\gamma }=0,$ see formulas
(\ref{deft}). This generates Finsler like solutions in Einstein gravity.

We emphasize that the system of Einstein--Finselr equations (\ref{eeqcdcc})
and constraints $Z_{\ \alpha \beta }^{\gamma }=0,$ for $\ ^{h}\lambda =$ $\
^{v}\lambda =\lambda ,$ is completely equivalent to (\ref{eeqlcc}) but it is
encoded more information about chosen nonholonomic frame structure in the
case with $\widehat{\mathbf{D}}.$ The surprising thing is that working in
general relativity with variables $(\mathbf{g},\widehat{\mathbf{D}})$ it is
more easy technically to construct exact solutions than using ''standard''
variables $(\mathbf{g},\nabla ).$ We can also extend the method for Finsler
theories and generalization. As a result, we were able to elaborate a new,
very general, geometric method of constructing exact solutions with generic
off--diagonal metrics and nontrivial nonholonomic structures in gravity
theories (the so--called anholonomic frame method, see reviews of results in
Refs. \cite{ijgmmp,vrflg,vsgg,vgsol}).

Nonholonomic configurations with $\ Z_{\ \alpha \beta }^{\gamma }=0$ present
an additional substantial interest because they allow us to model a subclass
of (pseudo) Finsler spaces as exact solutions in Einstein gravity. Such
Finsler metrics and connections are not restricted by modern experimental
data \cite{laem,will} and they do not involve violations of the local
Lorentz invariance, or metric noncompatibility of Finsler structures. In
next sections, we shall provide explicit examples of (pseudo) Finsler black
holes and nonholonomic configurations which can be modelled on Einstein
spaces by generic off--diagonal metrics.

\section{Off--Diagonal Ansatz and Exact Solutions}

We consider an ansatz for d--metric $\mathbf{g}$ (\ref{gpsm}) when the
\begin{equation}
\mathbf{g}=g_{i}dx^{i}\otimes dx^{i}+h_{3}\ {\delta v}\otimes {\delta v}%
+h_{4}\ {\delta y}\otimes {\delta y},\ \delta v=dv+w_{j}dx^{j},\ \delta
y=dy+n_{j}dx^{j}  \label{ans}
\end{equation}%
has nontrivial coefficients (smooth functions)
\begin{equation}
g_{i}=g_{i}(x^{k}),h_{a}=h_{a}(x^{i},v),w_{j}=w_{j}(x^{k},v),n_{j}=n_{j}(x^{k},v),
\label{coefdm}
\end{equation}%
where the N--connection coefficients are $N_{i^{\prime }}^{3}=w_{i^{\prime
}},$ $N_{i^{\prime }}^{4}=n_{i^{\prime }}$ and coordinates are parametrized
in the form $x^{k^{\prime }}=x^{k^{\prime }}(x^{k}),$ $y^{3^{\prime }}=v$
and $y^{4^{\prime }}=y.$ The above formulas are for a generic off--diagonal
metric with 2+2 splitting when the h--metric coefficients depend on two
variables and the v--metric and N--connection coefficients depend on three
variables $(x^{i},v)$. It has one Killing vector $e_{y}=\partial /\partial y$
because there is a frame basis when the coefficients do not depend on
variable $y.$

How to construct exact solutions for a general ansatz in general relativity
and fileld equations with nonholonomic variables in Einstein and generalized
Finsler gravity theories is analyzed in Refs. \cite{ijgmmp,vbe1,vbe2}, see
also generalizations and reviews of results in \cite{vrflg,vsgg,vncg}. In
this work, we shall omit technical details on constructing exact solutions
following the anholonomic frame method and send the reader to Ref. \cite%
{vgsol} providing general solutions of Einstein equations in arbitrary
dimensions, including both variants for equations (\ref{eeqcdcc}) and, in
particular, (\ref{eeqlcc}).

\subsection{Solutions for nonholonomic Einstein spaces, $\protect\lambda %
=const$}

A class of exact solutions of (\ref{eeqcdcc}) with cosmological constant for
the ansatz (\ref{ans}) is parametrized by d--metrics of type
\begin{eqnarray}
~~^{\lambda }\mathbf{\mathring{g}} &=&\epsilon _{1}e^{\underline{\phi }%
(x^{i})}\ dx^{1}\otimes dx^{1}+\epsilon _{2}e^{\underline{\phi }(x^{i})}\
dx^{2}\otimes dx^{2}  \label{lambsol} \\
&&+h_{3}\left( x^{k},v\right) \ ~{\delta v}\otimes ~{\delta v}+h_{4}\left(
x^{k},v\right) \ ~{\delta y}\otimes ~{\delta y},  \notag \\
~\delta v &=&dv+w_{i}\left( x^{k},v\right) dx^{i},\ ~\delta y=dy+n_{i}\left(
x^{k},v\right) dx^{i},  \notag
\end{eqnarray}%
for any signatures $\epsilon _{\alpha }=\pm 1,$ where the coefficients are
any functions satisfying (respectively) the conditions,%
\begin{eqnarray}
&&\epsilon _{1}\underline{\phi }^{\bullet \bullet }(x^{k})+\epsilon _{2}%
\underline{\phi }^{^{\prime \prime }}(x^{k})=-2\epsilon _{1}\epsilon _{2}\
\lambda ;  \label{coeflsoldc} \\
h_{3} &=&\pm \frac{\left( \phi ^{\ast }\right) ^{2}}{4\ \lambda }e^{-2\
^{0}\phi (x^{i})},\ h_{4}=\mp \frac{1}{4\ \lambda }e^{2(\phi -\ ^{0}\phi
(x^{i}))};\ w_{i}=-\partial _{i}\phi /\phi ^{\ast };  \notag \\
n_{i} &=&\ ^{1}n_{i}(x^{k})+\ ^{2}n_{i}(x^{k})\int \left( \phi ^{\ast
}\right) ^{2}e^{-2(\phi -\ ^{0}\phi (x^{i}))}dv,\   \notag \\
&=&\ \left\{
\begin{array}{rcl}
\ ^{1}n_{i}(x^{k})+\ ^{2}n_{i}(x^{k})\int e^{-4\phi }\frac{\left(
h_{4}^{\ast }\right) ^{2}}{h_{4}}dv,\ \  & \mbox{ if \ } & n_{i}^{\ast }\neq
0; \\
\ ^{1}n_{i}(x^{k}),\quad \qquad \qquad \qquad \qquad \qquad & \mbox{ if \ }
& n_{i}^{\ast }=0,%
\end{array}%
\right. \   \notag
\end{eqnarray}%
for any nonzero $h_{a}$ and $h_{a}^{\ast }$ and (integrating) functions $%
^{1}n_{i}(x^{k}),\ ^{2}n_{i}(x^{k}),$ generating function%
\begin{equation}
~\phi (x^{i},v)=\ln |h_{4}^{\ast }/\sqrt{|h_{3}h_{4}|}|  \label{auxphi}
\end{equation}%
and $\ ^{0}\phi (x^{i})$ to be determined from certain boundary conditions
for a fixed system of coordinates. In above formulas, we consider partial
derivatives written in the form $a^{\bullet }=\partial a/\partial x^{1},$\ $%
a^{\prime }=\partial a/\partial x^{2},$\ $a^{\ast }=\partial a/\partial v.$
There are two classes of solutions (\ref{coeflsoldc}) constructed for a
nontrivial $\lambda .$ The first one is singular for $\lambda \rightarrow 0$
if we chose a generation function $\phi (x^{i},v)$ (\ref{auxphi}) \ not
depending on $\lambda .$ We can eliminate such singularities for parametric
dependencies of type $\phi (\lambda ,x^{i},v),$ when such a function is
linear on $\lambda .$

We have to impose additional constraints on coefficients of a chosen ansatz
in order to satisfy the conditions $Z_{\ \alpha \beta }^{\gamma }=0$ (\ref%
{deft}) and generate solutions of the Einstein equations (\ref{eeqlcc}) for
the Levi--Civita connection:
\begin{eqnarray}
(2e^{2\phi }\phi -\lambda )\left( \phi ^{\ast }\right) ^{2} &=&0,\phi \neq
0,\phi ^{\ast }\neq 0;  \label{lcls} \\
w_{1}w_{2}\left( \ln |\frac{w_{1}}{w_{2}}|\right) ^{\ast } &=&w_{2}^{\bullet
}-w_{1}^{\prime },w_{i}^{\ast }\neq 0;\ w_{2}^{\bullet }-w_{1}^{\prime }
=0,\ w_{i}^{\ast }=0;  \notag \\
\ ^{1}n_{1}^{\prime }(x^{k})-\ ^{1}n_{2}^{\bullet }(x^{k}) &=&0,\
n_{i}^{\ast }=0;  \notag
\end{eqnarray}%
we can consider $\phi (x^{i},v)=const$ if we include configurations with $%
\phi ^{\ast }=0.$

\subsection{Solutions for (non) holonomic vacuum spaces, $\protect\lambda =0$%
}

Because of generic nonlinear off--diagonal and possible nonholonomic
character of solutions of Einstein equations, the vacuum solutions are
generated not just as a simple limit $\lambda \rightarrow 0$ of coefficients
(\ref{coeflsoldc}) and (\ref{lcls}). Such a limit to vacuum configurations
should be considered for the gravitational field equations with zero sources
on the right part with a further integration on separated variables, see
details in \cite{ijgmmp,vrflg,vsgg,vncg}. This way we construct a class of
vacuum solutions of the Einstein equations for the canonical d--connection, $%
\widehat{\mathbf{R}}_{\alpha \beta }=0$ in (\ref{eeqcdcc}) by d-metrics $~%
\mathbf{\mathring{g}}$ parametrized in the form (\ref{lambsol}) but with
coefficients satisfying conditions%
\begin{eqnarray}
&&\epsilon _{1}\underline{\phi }^{\bullet \bullet }(x^{k})+\epsilon _{2}%
\underline{\phi }^{^{\prime \prime }}(x^{k})=0;  \label{vacdsolc} \\
h_{3} &=&\pm e^{-2\ ^{0}\phi }\frac{\left( h_{4}^{\ast }\right) ^{2}}{h_{4}}%
\mbox{ for  given }h_{4}(x^{i},v),\ \phi =\ ^{0}\phi =const;\   \notag \\
w_{i} &=&w_{i}(x^{i},v),\mbox{ for any such functions if }\lambda =0;  \notag
\\
n_{i} &=&\ \left\{
\begin{array}{rcl}
\ \ \ ^{1}n_{i}(x^{k})+\ ^{2}n_{i}(x^{k})\int \left( h_{4}^{\ast }\right)
^{2}|h_{4}|^{-5/2}dv,\ \  & \mbox{ if \ } & n_{i}^{\ast }\neq 0; \\
\ ^{1}n_{i}(x^{k}),\quad \qquad \qquad \qquad \qquad \qquad & \mbox{ if \ }
& n_{i}^{\ast }=0.%
\end{array}%
\right.  \notag
\end{eqnarray}

We get vacuum solutions $~_{\shortmid }\mathbf{\mathring{g}}$ of the
Einstein equations (\ref{eeqlcc}) for the Levi--Civita connection, i.e of $%
R_{\alpha \beta }=0,$ if we impose additional constraints on coefficients of
d--metric, for $e^{-2\ ^{0}\phi }=1,$
\begin{eqnarray}
h_{3} &=&\pm 4\left[ \left( \sqrt{|h_{4}|}\right) ^{\ast }\right] ^{2},\quad
h_{4}^{\ast }\neq 0;  \notag \\
w_{1}w_{2}\left( \ln |\frac{w_{1}}{w_{2}}|\right) ^{\ast } &=&w_{2}^{\bullet
}-w_{1}^{\prime },\ w_{i}^{\ast }\neq 0;\ w_{2}^{\bullet }-w_{1}^{\prime }
=0,\ w_{i}^{\ast }=0;  \notag \\
\ ^{1}n_{1}^{\prime }(x^{k})-\ ^{1}n_{2}^{\bullet }(x^{k}) &=&0,\
n_{i}^{\ast }=0.  \label{vaclcsoc}
\end{eqnarray}

It should be emphasized that the bulk of vacuum and cosmological solutions
in general relativity outlined in Refs. \cite{kramer,bic} can be considered
as particular cases of metrics with $h_{4}^{\ast }=0,w_{i}^{\ast }=0$ and/or
$n_{i}^{\ast }=0,$ for corresponding systems of reference. In our approach,
we work with more general classes of off--diagonal metrics with certain
coefficients depending on three variables. Such solutions in general
relativity can be generated if we impose certain nonholonomic constraints on
integral varieties of corresponding systems of partial equations. The former
analytic and computer numeric programs (for instance, the standard ones with
Maple/ Mathematica) for constructing solutions in gravity theories can not
be directly applied for alternative verifications of our solutions because
those approaches do not encode constraints of type (\ref{lcls}) or (\ref%
{vaclcsoc}). Nevertheless, it is possible to check in general analytic form,
see all details summarized in Part II of \cite{vsgg} and \cite{vgsol}, that
the Einstein equations are satisfied for such general off--diagonal ansatz
of metrics and various types of d-- and N--connections.

\subsection{Nonholonomic deformations of the Schwarzschild metric}

\label{sschw}We consider a diagonal metric
\begin{equation}
~^{\varepsilon }\mathbf{g}=-d\xi \otimes d\xi -r^{2}(\xi )\ d\vartheta
\otimes d\vartheta -r^{2}(\xi )\sin ^{2}\vartheta \ d\varphi \otimes
d\varphi +\varpi ^{2}(\xi )\ dt\otimes \ dt,  \label{5aux1}
\end{equation}%
where the local coordinates and nontrivial metric coefficients are
parametriz\-ed in the form%
\begin{equation}
\check{g}_{1} = -1,\ \check{g}_{2}=-r^{2}(\xi ),\ \check{h}_{3}=-r^{2}(\xi
)\sin ^{2}\vartheta ,\ \check{h}_{4}=\varpi ^{2}(\xi ),  \label{5aux1p}
\end{equation}
for $x^{1} =\xi ,x^{2}=\vartheta ,y^{3}=\varphi ,y^{4}=t,$ where $\xi =\int
dr\ \left| 1-\frac{2\mu _{0}}{r}+\frac{\varepsilon }{r^{2}}\right| ^{1/2}$
and $\varpi ^{2}(r)=1-\frac{2\mu _{0}}{r}+\frac{\varepsilon }{r^{2}}.$ For
the constants $\varepsilon = 0$ and $\mu _{0}$ being a point mass, the
metric $~^{\varepsilon }\mathbf{g}$ (\ref{5aux1}) is just that for the
Schwarzschild solution written in spacetime spherical coordinates $%
(r,\vartheta ,\varphi ,t).$\footnote{%
For simplicity, in this work, we \ shall consider only the case of ''pure''
gravitational vacuum solutions, not analyzing a more general possibility
when $\varepsilon =e^{2}$ can be related to the electric charge for the
Reissner--Nordstr\"{o}m metric for the so--called holonomic electro vacuum
configurations(see details, for example, \cite{heu}). We treat $\varepsilon $
as a small parameter (eccentricity) defining a small deformation of a circle
into an ellipse.}

Let us consider nonholonomic deformations when $g_{i}=\eta _{i}\check{g}_{i}$
and $h_{a}=\eta _{a}\check{h}_{a}$ and $w_{i},n_{i}$ are some nontrivial
functions, where $(\check{g}_{i},\check{h}_{a})$ are given by data (\ref%
{5aux1p}), to an ansatz
\begin{eqnarray}
~_{\eta }^{\varepsilon }\mathbf{g} &=&-\eta _{1}(\xi )d\xi \otimes d\xi
-\eta _{2}(\xi )r^{2}(\xi )\ d\vartheta \otimes d\vartheta  \label{5sol1} \\
&&-\eta _{3}(\xi ,\vartheta ,\varphi )r^{2}(\xi )\sin ^{2}\vartheta \ \delta
\varphi \otimes \delta \varphi +\eta _{4}(\xi ,\vartheta ,\varphi )\varpi
^{2}(\xi )\ \delta t\otimes \delta t,  \notag \\
\delta \varphi &=&d\varphi +w_{1}(\xi ,\vartheta ,\varphi )d\xi +w_{2}(\xi
,\vartheta ,\varphi )d\vartheta ,\   \notag \\
\delta t &=&dt+n_{1}(\xi ,\vartheta )d\xi +n_{2}(\xi ,\vartheta )d\vartheta ,
\notag
\end{eqnarray}%
for which the coefficients are constrained to define nonholonomic Einstein
spaces when the conditions (\ref{lambsol}) are satisfied. There are used
3--d spacial spherical coordinates $(\xi (r),\vartheta ,\varphi ),$ or $%
(r,\vartheta ,\varphi ),$ for a class of metrics of type (\ref{ans}) with
coefficients of type (\ref{coefdm}).

The equations (\ref{eeqcdcc}) for zero source state certain relations
between the coefficients of the vertical metric and respective polarization
functions,\footnote{%
for details, we send readers to a proof of Theorem 4.1 in Ref. \cite{vgsol}}%
\begin{equation}
h_{3} =-h_{0}^{2}(b^{\ast })^{2}=\eta _{3}(\xi ,\vartheta ,\varphi
)r^{2}(\xi )\sin ^{2}\vartheta ,\ h_{4} =b^{2}=\eta _{4}(\xi ,\vartheta
,\varphi )\varpi ^{2}(\xi ),  \label{aux41}
\end{equation}%
for $|\eta _{3}|=(h_{0})^{2}|\check{h}_{4}/\check{h}_{3}|\left[ \left( \sqrt{%
|\eta _{4}|}\right) ^{\ast }\right] ^{2}.$ In these formulas, we have to
chose $h_{0}=const$ \ (it must be $h_{0}=2$ in order to satisfy the first
condition (\ref{vaclcsoc})), where $\check{h}_{a}$ are stated by the
Schwarzschild solution for the chosen system of coordinates and $\eta _{4}$
can be any function satisfying the condition $\eta _{4}^{\ast }\neq 0.$ We
generate a class of solutions for any function $b(\xi ,\vartheta ,\varphi )$
with $b^{\ast }\neq 0.$ For different purposes, it is more convenient to
work directly with $\eta _{4},$ for $\eta _{4}^{\ast }\neq 0,$ and/or $%
h_{4}, $ for $h_{4}^{\ast }\neq 0.$

It is possible to compute the polarizations $\eta _{1}$ and $\eta _{2},$
when $\eta _{1}=\eta _{2}r^{2}=e^{\psi (\xi ,\vartheta )},$ from (\ref%
{eeqcdcc}) with zero source, written in the form $\psi ^{\bullet \bullet
}+\psi ^{\prime \prime }=0.$

Introducing the defined values of the coefficients in the ansatz (\ref{5sol1}%
), we find a class of exact vacuum solutions of the Einstein equations
defining stationary nonholonomic deformations of the Sch\-warz\-schild
metric,
{\small
\begin{eqnarray}
~^{\varepsilon }\mathbf{g} &=&-e^{\psi }\left( d\xi \otimes d\xi +\
d\vartheta \otimes d\vartheta \right)   -4\left[ \left( \sqrt{|\eta _{4}|}\right) ^{\ast }\right] ^{2}\varpi ^{2}\
\delta \varphi \otimes \ \delta \varphi +\eta _{4}\varpi ^{2}\ \delta
t\otimes \delta t, \notag   \\
\delta \varphi &=&d\varphi +w_{1}(\xi ,\vartheta ,\varphi )d\xi +w_{2}(\xi
,\vartheta ,\varphi )d\vartheta ,\ \delta t =dt+\ ^{1}n_{1}d\xi +\
^{1}n_{2}d\vartheta .  \label{5sol1a} 
\end{eqnarray}%
}
The N--connection coefficients $w_{i}$ and $\ ^{1}n_{i}$ must satisfy the
conditions (\ref{vaclcsoc}) in order to get vacuum metrics in Einstein
gravity. Such vacuum solutions are for nonholonomic deformations of a static
black hole metric into (non) holonomic Einstein spaces with locally
anistoropic backgrounds (on coordinate $\varphi )$ defined by an arbitrary
function $\eta _{4}(\xi ,\vartheta ,\varphi )$ with $\partial _{\varphi
}\eta _{4}\neq 0,$ an arbitrary $\psi (\xi ,\vartheta )$ solving the 2--d
Laplace equation and certain integration functions $\ ^{1}w_{i}(\xi
,\vartheta,\varphi )$ and $\ ^{1}n_{i}(\xi ,\vartheta ).$

In general, the solutions from the target set of metrics do not define black
holes and do not describe obvious physical situations. Nevertheless, they
preserve the singular character of the coefficient $\varpi ^{2}$ vanishing
on the horizon of a Schwarzschild black hole if we take only smooth
integration functions for some small deformation parameters $\varepsilon .$
We can also consider a prescribed physical situation when, for instance, $%
\eta _{4}$ mimics 3--d, or 2--d, solitonic polarizations on coordinates $\xi
,\vartheta ,\varphi ,$ or on $\xi ,\varphi .$

\subsection{Solutions with linear parametric nonholonomic polarizations}

\label{ssvsol}For a general d--metric (\ref{5sol1a}) defining nonholonomic
deformations of the Schwarzschild solution depending on parameter $%
\varepsilon ,$ we can select locally anisotropic configurations with
possible physical interpretation of gravitational vacuum configurations with
spherical and/or rotoid (ellipsoid) symmetry. Let us consider a generating
function of type
\begin{equation}
b^{2}=q(\xi ,\vartheta ,\varphi )+\varepsilon s(\xi ,\vartheta ,\varphi )
\label{gf1}
\end{equation}%
and, for simplicity, restrict our analysis only with linear decompositions
on a small parameter $\varepsilon ,$ with $0<\varepsilon <<1.$ This way, we
shall construct exact solutions with off--diagonal metrics of the\ Einstein
equations depending on $\varepsilon $ which for rotoid configurations can be
considered as a small eccentricity. From a formal point of view, we can
summarize on all orders $\varepsilon ^{2},$ $\varepsilon ^{3}...$ stating
such recurrent formulas for coefficients when get convergent series to some
functions depending both on spacetime coordinates and a parameter $%
\varepsilon ,$ see a detailed analysis in Ref. \cite{ijgmmp}.

A straightforward computation with (\ref{gf1}) allows us to write $\ \left(
b^{\ast }\right) ^{2}=[(\sqrt{|q|})^{\ast }]^{2}[1+\varepsilon \frac{1}{(%
\sqrt{|q|})^{\ast }}(\frac{s}{\sqrt{|q|}})^{\ast }]$ and compute the
vertical coefficients of d--metric (\ref{5sol1a}), i.e $h_{3}$ and $h_{4}$
(and corresponding polarizations $\eta _{3}$ and $\eta _{4})$ using formulas
(\ref{aux41}). In a particular case, we can generate nonholonomic
deformations of the Schwarzschild solution not depending on $\varepsilon $
if we consider $\varepsilon =0$ in the above formulas consider only
nonholonomic deformations with $b^{2}=q$ and $\left( b^{\ast }\right) ^{2}=%
\left[ (\sqrt{|q|})^{\ast }\right] ^{2}.$

Nonholonomic deformations to rotoid configurations are possible if we chose
\begin{equation}
q=1-\frac{2\mu (\xi ,\vartheta ,\varphi )}{r}\mbox{ and }s=\frac{q_{0}(r)}{%
4\mu ^{2}}\sin (\omega _{0}\varphi +\varphi _{0}),  \label{aux42}
\end{equation}%
for $\mu (\xi ,\vartheta ,\varphi )=\mu _{0}+\varepsilon \mu _{1}(\xi
,\vartheta ,\varphi )$ (locally anisotropically polarized mass) with certain
constants $\mu ,\omega _{0}$ and $\varphi _{0}$ and arbitrary
functions/polarizations $\mu _{1}(\xi ,\vartheta ,\varphi )$ and $q_{0}(r)$
to be determined from some boundary conditions, with $\varepsilon $ being
the eccentricity.\footnote{%
A nonholonomic Einstein vacuum can be modelled as a continuum media with
possible singularities which because of generic nonlinear character of
gravitational interactions may result in effective locally anisotropic
polarizations of fundamental physical constants. Similar effects of
polarization of constants can be measured experimentally in classical and
nonlinear electrodynamics, for instance, in various types of continuous
media and dislocations and disclinations. For spherical symmetries, with
local isotropy, the point mass is approximated by a constant $\mu _{0}.$ We
have to consider anisotropically polarized masses of type $\mu _{1}(\xi
,\varphi ,\vartheta )$ for locally anisotropic models (for instance, with
ellipsoidal symmetry) in general relativity and various types of gravity
theories. Such sources should be introduced following certain
phenomenological arguments, like in classical electrodynamics, or computed
as certain quasi--classical approximations from a quantum gravity model,
like in quantum electrodynamics.} The possibility to treat $\varepsilon $ as
an eccentricity follows from the condition that the coefficient $%
h_{4}=b^{2}=\eta _{4}(\xi ,\vartheta ,\varphi )\varpi ^{2}(\xi )$ becomes
zero for data (\ref{aux42}) if $r_{+}\simeq 2\mu _{0}/\left( 1+\varepsilon
\frac{q_{0}(r)}{4\mu ^{2}}\sin (\omega _{0}\varphi +\varphi _{0})\right) .$
This condition defines a small deformation of the Schwarzschild spherical
horizon into an ellipsoidal one (rotoid configuration with eccentricity $%
\varepsilon ).$

Let us summarize the results for d--metrics defining rotoid type solutions:
\begin{eqnarray}
~^{rot}\mathbf{g} &=&-e^{\psi }\left( d\xi \otimes d\xi +\ d\vartheta
\otimes d\vartheta \right) +\left( q+\varepsilon s\right) \ \delta t\otimes
\delta t  \notag \\
&&-4\left[ (\sqrt{|q|})^{\ast }\right] ^{2}[1+\varepsilon \frac{1}{(\sqrt{|q|%
})^{\ast }}(\frac{s}{\sqrt{|q|}})^{\ast }]\ \delta \varphi \otimes \ \delta
\varphi ,,  \label{rotoidm} \\
\delta \varphi  &=&d\varphi +w_{1}d\xi +w_{2}d\vartheta ,\ \delta t=dt+\
^{1}n_{1}d\xi +\ ^{1}n_{2}d\vartheta ,  \notag
\end{eqnarray}%
with functions $q(\xi ,\vartheta ,\varphi )$ and $s(\xi ,\vartheta ,\varphi )
$ given by formulas (\ref{aux42}) and N--connec\-ti\-on coefficients $%
w_{i}(\xi ,\vartheta ,\varphi )$ and $\ n_{i}=$ $\ ^{1}n_{i}(\xi ,\vartheta )
$ subjected to conditions of type (\ref{vaclcsoc}),
\begin{eqnarray*}
w_{1}w_{2}\left( \ln |\frac{w_{1}}{w_{2}}|\right) ^{\ast } &=&w_{2}^{\bullet
}-w_{1}^{\prime },\quad w_{i}^{\ast }\neq 0; \\
\mbox{ or \ }w_{2}^{\bullet }-w_{1}^{\prime } &=&0,\quad w_{i}^{\ast }=0;\
^{1}n_{1}^{\prime }(\xi ,\vartheta )-\ ^{1}n_{2}^{\bullet }(\xi ,\vartheta
)=0
\end{eqnarray*}%
and $\psi (\xi ,\vartheta )$ being any function for which $\psi ^{\bullet
\bullet }+\psi ^{\prime \prime }=0.$

Finally, we emphasize that the d--metrics with rotoid symmetry constructed
in this section are different from those considered in our previous works
\cite{vbe1,vbe2,vncg}. In general, they do not define black hole solutions.
Nevertheless, for small eccentricities, we get stationary configurations for
the so--called black ellipsoid solutions (their stability and properties can
be analyzed following the methods elaborated in the mentioned works, see
also a summary of results and generalizations for various types of locally
anisotropic gravity models in Ref. \cite{vsgg}).

\section{Finsler Black Holes, Ellipsoids and Nonlinear Gravitational Waves}

The next step to be taken is to show how we can construct black hole
solutions in a (pseudo) Finsler spacetime using certain analogy with the
Schwarzschild solution rewritten in Finsler variables. There are several
avenues to be explored, and we separate the material into three subsections.
The first one is for nonholonomic rotoid deformations of Einstein metrics
when the resulting general off--diagonal metrics contain a nontrivial
cosmological constant. The second one concerns imbedding of black hole
solutions and their nonholonomic deformations into nontrivial backgrounds of
nonlinear waves. Finally, the third subsection is devoted to Finsler
variables in general relativity and analogs of the Schwarzschild solution in
(pseudo) Finsler spacetimes.

\subsection{Nonholonomic rotoid deformations of Einstein metrics}

Using the anholonomic frame method, we can construct a class of solutions
with nontrivial cosmological constant possessing different limits, for large
radial distances and small nonholonomic deformations, than vacuum
configurations considered in section \ref{ssvsol}. Such stationary metrics
belong to the class of d--metrics (\ref{lambsol}) defining exact solutions
of gravitational equations (\ref{eeqcdcc}).

Let us consider a diagonal metric of type
\begin{equation}
~_{\lambda }^{\varepsilon }\mathbf{g}=-d\xi \otimes d\xi -r^{2}(\xi )\
d\vartheta \otimes d\vartheta -r^{2}(\xi )\sin ^{2}\vartheta \ d\varphi
\otimes d\varphi +\ _{\lambda }\varpi ^{2}(\xi )\ dt\otimes \ dt,
\label{sds1}
\end{equation}%
where the local coordinates and nontrivial metric coefficients are
parametriz\-ed in the form%
\begin{equation*}
\check{g}_{1}=-1,\ \check{g}_{2}=-r^{2}(\xi ),\ \check{h}_{3}=-r^{2}(\xi
)\sin ^{2}\vartheta ,\ \check{h}_{4}=\ _{\lambda }\varpi ^{2}(\xi )
\end{equation*}%
where $x^{1}=\xi ,x^{2}=\vartheta ,y^{3}=\varphi ,y^{4}=t,$ for $\xi =\int \
\left| 1-\frac{2\underline{\mu }}{r}+\varepsilon (\frac{1}{r^{2}}+\frac{\
\underline{\lambda }}{3}\ _{4}\kappa ^{2}\ r^{2})\right| ^{\frac{1}{2}}$ $%
dr, $ and $\ _{\lambda }\varpi ^{2}(r)=1-\frac{2\underline{\mu }}{r}%
+\varepsilon \left( \frac{1}{r^{2}}-\frac{\ \underline{\lambda }}{3}\
_{4}\kappa ^{2}\ r^{2}\right) ;$ $\ _{4}\kappa ^{2}=1/M_{\ast }^{2}$ stands
for the 4--dimensional Newton's constant, $\lambda =$ $\varepsilon \ \
\underline{\lambda }$ \ is a positive cosmological constant and $\mu _{1}$
is the so--called ADM mass, see Ref. \cite{kanti} for a review of results on
Schwarzschild--de Sitter black holes in $(4+n_{1})$--dimensions, for $%
n_{1}=1,2,...$ For the constants $\varepsilon \rightarrow 0$ and $\underline{%
\mu }$ taken to be a point mass (in general, for a stationary locally \
anisotropic model this is a function of type \ $\underline{\mu }=$\ $%
\underline{\mu }_{0}+\varepsilon $\ $\underline{\mu }_{1}\left( \xi
,\vartheta ,\varphi \right) ,$ for \ $\underline{\mu }_{0}=const$ and
function $\underline{\mu }_{1}\left( \xi ,\vartheta ,\varphi \right) $ taken
from phenomenological considerations), the metric $~^{\varepsilon }\mathbf{g}
$ \ (\ref{sds1}) has a true singularity at $r=0.$ The equation $1-\frac{2%
\underline{\mu }_{0}}{r}+\frac{1}{3}\lambda \ _{4}\kappa ^{2}\ r^{2}=0$ has
three solutions for not small $r$ (when we can neglect the term $1/r^{2})$
corresponding to three horizons for this spacetime. There are only two real
positive roots because of positivity of radial coordinate $r:$ \ the first
one corresponds to the so--called ''cosmological horizon'' and the second
one (the smaller) is for the ''black hole event horizon''. A nontrivial
parameter $\varepsilon $ deforms the metric black hole metric
nonholonomically into a d--metric which (in general) does not satisfy the
Einstein equations. We have to introduce additional off--diagonal terms and
new nonholonomic constraints in order to define nonholonomic transforms into
an exact solution.

We chose local coordinates as in (\ref{5aux1p}) and consider the ansatz
{\small 
\begin{eqnarray*}
~~^{\lambda }\mathbf{\mathring{g}} &=&-e^{\underline{\phi }(\xi ,\vartheta
)}\ d\xi \otimes d\xi -e^{\underline{\phi }(\xi ,\vartheta )}\ d\vartheta
\otimes d\vartheta +h_{3}\left( \xi ,\vartheta ,\varphi \right) \ ~{\delta }\varphi \otimes ~{%
\delta }\varphi +h_{4}\left( \xi ,\vartheta ,\varphi \right) \ ~{\delta t}%
\otimes ~{\delta t}, \\
~\delta \varphi &=&d\varphi +w_{1}\left( \xi ,\vartheta ,\varphi \right)
d\xi +w_{2}\left( \xi ,\vartheta ,\varphi \right) d\vartheta , \\
\ \delta t &=&dt+n_{1}\left( \xi ,\vartheta ,\varphi \right) d\xi
+n_{2}\left( \xi ,\vartheta ,\varphi \right) d\vartheta ,
\end{eqnarray*}%
}
for $h_{3}=-h_{0}^{2}(b^{\ast })^{2}=\eta _{3}(\xi ,\vartheta ,\varphi
)r^{2}(\xi )\sin ^{2}\vartheta ,\ h_{4}=b^{2}=\eta _{4}(\xi ,\vartheta
,\varphi )\ _{\lambda }\varpi ^{2}(\xi ),$ where the coefficients satisfy
the conditions,%
\begin{eqnarray}
&&\underline{\phi }^{\bullet \bullet }(\xi ,\vartheta )+\underline{\phi }%
^{^{\prime \prime }}(\xi ,\vartheta )=2\ \lambda ;  \label{anhsol2} \\
h_{3} &=&\pm \frac{\left( \phi ^{\ast }\right) ^{2}}{4\ \lambda }e^{-2\
^{0}\phi (\xi ,\vartheta )},\ h_{4}=\mp \frac{1}{4\ \lambda }e^{2(\phi -\
^{0}\phi (\xi ,\vartheta ))};\ w_{i}=-\partial _{i}\phi /\phi ^{\ast };
\notag \\
n_{i} &=&\ ^{1}n_{i}(\xi ,\vartheta )+\ ^{2}n_{i}(\xi ,\vartheta )\int
\left( \phi ^{\ast }\right) ^{2}e^{-2(\phi -\ ^{0}\phi (\xi ,\vartheta
))}d\varphi ,\   \notag \\
&=&\ \ \left\{
\begin{array}{rcl}
\ \ ^{1}n_{i}(\xi ,\vartheta )+\ ^{2}n_{i}(\xi ,\vartheta )\int e^{-4\phi }%
\frac{\left( h_{4}^{\ast }\right) ^{2}}{h_{4}}d\varphi ,\ \  & \mbox{ if \ }
& n_{i}^{\ast }\neq 0; \\
\ ^{1}n_{i}(\xi ,\vartheta ),\quad \qquad \qquad \qquad \qquad \qquad & %
\mbox{ if \ } & n_{i}^{\ast }=0;%
\end{array}%
\right.  \notag
\end{eqnarray}%
for any nonzero $h_{a}$ and $h_{a}^{\ast }$ and (integrating) functions $%
^{1}n_{i}(\xi ,\vartheta ),\ ^{2}n_{i}(\xi ,\vartheta ),$ generating
function $\phi (\xi ,\vartheta ,\varphi )$ (\ref{auxphi}), and $\ ^{0}\phi
(\xi ,\vartheta )$ to be determined from certain boundary conditions for a
fixed system of coordinates.

In explicit form, the d--metric determining nonholonomic ellipsoid de Sitter
configurations is written
\begin{eqnarray}
~_{\lambda }^{rot}\mathbf{g} &=&-e^{\underline{\phi }(\xi ,\vartheta
)}\left( d\xi \otimes d\xi +\ d\vartheta \otimes d\vartheta \right) +\left(
\underline{q}+\varepsilon \underline{s}\right) \ \delta t\otimes \delta t
\notag \\
&&-h_{0}^{2}\left[ (\sqrt{|\underline{q}|})^{\ast }\right] ^{2}\left[
1+\varepsilon \frac{1}{(\sqrt{|\underline{q}|})^{\ast }}\left( \frac{%
\underline{s}}{\sqrt{|\underline{q}|}}\right) ^{\ast }\right] \ \delta
\varphi \otimes \ \delta \varphi ,  \notag \\
\delta \varphi &=&d\varphi +w_{1}d\xi +w_{2}d\vartheta ,\ \delta
t=dt+n_{1}d\xi +n_{2}d\vartheta ,  \label{soladel}
\end{eqnarray}%
where
\begin{equation*}
\underline{q}=1-\frac{2\ ^{1}\underline{\mu }(r,\vartheta ,\varphi )}{r},\ \
\underline{s}=\frac{\underline{q}_{0}(r)}{4\underline{\mu }_{0}^{2}}\sin
(\omega _{0}\varphi +\varphi _{0}),
\end{equation*}%
with $\ ^{1}\underline{\mu }(r,\vartheta ,\varphi )=\underline{\mu }%
+\varepsilon \left( r^{-2}-\underline{\lambda }\ _{4}\kappa ^{2}\
r^{2}/3\right) /2,$ chosen to generate an anisotro\-pic rotoid configuration
for the smaller ''horizon'' (when $\ h_{4}=0),\ r_{+}\simeq 2\ ^{1}%
\underline{\mu }/\left( 1+\varepsilon \frac{\underline{q}_{0}(r)}{4%
\underline{\mu }_{0}^{2}}\sin (\omega _{0}\varphi +\varphi _{0})\right) ,$
for a corresponding $\underline{q}_{0}(r).$ The d--metric (\ref{soladel}) \
and N--connection coefficients (\ref{anhsol2}) determines a solution of
nonholonomic Einstein equations (\ref{eeqcdcc}). It is not a solution in
general relativity but can be considered in (pseudo) Finsler models of
gravity.

We have to impose the condition that the coefficients of the above d--metric
satisfy the constraints \ (\ref{lcls}) in order to generate solutions of the
Einstein equations (\ref{eeqlcc}) for the Levi--Civita connection. From the
first constraint, for $\phi ^{\ast }\neq 0,$ we obtain the condition that $%
\phi (r,\varphi ,\vartheta )=\ln |h_{4}^{\ast }/\sqrt{|h_{3}h_{4}|}|$ must
be any function defined in non--explicit form from equation $2e^{2\phi }\phi
=\lambda .$ The set of constraints for the N--connection coefficients is to
be satisfied if the integration functions in (\ref{anhsol2}) are chosen in a
form when $w_{1}w_{2}\left( \ln |\frac{w_{1}}{w_{2}}|\right) ^{\ast
}=w_{2}^{\bullet }-w_{1}^{\prime }$ for $w_{i}^{\ast }\neq 0;$ $%
w_{2}^{\bullet }-w_{1}^{\prime }=0$ for$\ w_{i}^{\ast }=0;$ and take $\
n_{i}=\ ^{1}n_{i}(x^{k})$ for $\ ^{1}n_{1}^{\prime }(x^{k})-\
^{1}n_{2}^{\bullet }(x^{k})=0.$

In the limit $\varepsilon \rightarrow 0,$ we get a subclass of solutions of
type (\ref{soladel}) possessing spherical symmetry but with generic
off--diagonal coefficients induced by the N--connection coefficients and
depending on cosmological constant. In order to extract from such
configurations the Schwarzschild solution, we must select a set of functions
with the properties $\phi \rightarrow const,w_{i}\rightarrow
0,n_{i}\rightarrow 0$ and $h_{4}\rightarrow \varpi ^{2}.$ In general, the
parametric dependence on cosmological constant, for nonholonomic
configurations, is not smooth.

\subsection{Rotoids and solitonic distributions}

On a N--anholonomic spacetime $\mathbf{V}$ defined by a rotoid d--metric $%
~^{rot}\mathbf{g}$ (\ref{rotoidm}), we can consider a static three
dimensional solitonic distribution $\eta (\xi ,\vartheta ,\varphi )$ as a
solution of solitonic equation\footnote{%
as a matter of principle, we can consider\ that $\eta $ is a solution of any
three dimensional solitonic and/ or other nonlinear wave equations}
\begin{equation*}
\eta ^{\bullet \bullet }+\epsilon (\eta ^{\prime }+6\eta \ \eta ^{\ast
}+\eta ^{\ast \ast \ast })^{\ast }=0,\ \epsilon =\pm 1.
\end{equation*}%
It is possible to define a nonholonomic transform from $^{rot}\mathbf{g}$ to
a d--metric $~_{st}^{rot}\mathbf{g}$ determining a stationary metric for a
rotoid in solitonic background in general relativity:
\begin{eqnarray}
~_{st}^{rot}\mathbf{g} &=&-e^{\psi }\left( d\xi \otimes d\xi +\ d\vartheta
\otimes d\vartheta \right) +\eta \left( q+\varepsilon s\right) \ \delta
t\otimes \delta t  \label{solrot} \\
&&-4\left[ (\sqrt{|\eta q|})^{\ast }\right] ^{2}[1+\varepsilon \frac{1}{(%
\sqrt{|\eta q|})^{\ast }}(\frac{s}{\sqrt{|\eta q|}})^{\ast }]\ \delta
\varphi \otimes \ \delta \varphi ,,  \notag \\
\delta \varphi  &=&d\varphi +w_{1}d\xi +w_{2}d\vartheta ,\ \delta t=dt+\
^{1}n_{1}d\xi +\ ^{1}n_{2}d\vartheta ,  \notag
\end{eqnarray}%
where the N--connection coefficients are taken the same as for (\ref{rotoidm}%
). In the limit $\varepsilon \rightarrow 0,$ this metric defines a
nonholonomic imbedding of the Schwarzschild solution into a solitonic
vacuum, which results in a vacuum solution of the Einstein gravity defined
by a stationary generic off--diagonal metric. For small polarizations, when $%
|\eta |\sim 1,$ it is preserved the black hole character of metric and the
solitonic distribution can be considered as on a Schwarzschild background.
It is also possible to take such parameters of $\eta $ when a black hole is
nonholonomically placed on a \textquotedblright gravitational
hill\textquotedblright\ defined by a soliton.

A d--metric (\ref{solrot}) can be generalized for (pseudo) Finsler spaces
with canonical d--connection as a solution of equations $\widehat{\mathbf{R}}%
_{\alpha \beta }=0$ (\ref{eeqcdcc}) by d-metrics parametrized in the form (%
\ref{lambsol}) with stationary coefficients subjected to conditions
\begin{eqnarray}
&&\psi ^{\bullet \bullet }(\xi ,\vartheta )+\psi ^{^{\prime \prime }}(\xi
,\vartheta )=0;  \label{aux2a} \\
h_{3} &=&\pm e^{-2\ ^{0}\phi }\frac{\left( h_{4}^{\ast }\right) ^{2}}{h_{4}}%
\mbox{ for  given }h_{4}(\xi ,\vartheta ,\varphi ),\ \phi =\ ^{0}\phi
=const;\   \notag \\
w_{i} &=&w_{i}(\xi ,\vartheta ,\varphi )\mbox{ are any functions if }\lambda
=0;  \notag \\
n_{i} &=&\ \ \left\{
\begin{array}{rcl}
\ ^{1}n_{i}(\xi ,\vartheta )+\ ^{2}n_{i}(\xi ,\vartheta )\int \left(
h_{4}^{\ast }\right) ^{2}|h_{4}|^{-5/2}dv,\  & \mbox{ if \ } & n_{i}^{\ast
}\neq 0; \\
\ ^{1}n_{i}(\xi ,\vartheta ),\quad \qquad \qquad \qquad \qquad \qquad & %
\mbox{ if \ } & n_{i}^{\ast }=0;%
\end{array}%
\right.  \notag
\end{eqnarray}%
for $h_{4}=\eta (\xi ,\vartheta ,\varphi )\left[ q(\xi ,\vartheta ,\varphi
)+\varepsilon s(\xi ,\vartheta ,\varphi )\right] .$ In the limit $%
\varepsilon \rightarrow 0,$ we get a so--called Schwarzschild black hole
solution mapped nonholonomically on a N--anholonomic (pseudo) Riemannian
spacetime, or on a nonholonomic tangent bundle.

We get a model of Finsler gravity on a tangent bundle $TM$ with a
two--dimensional base $M$ and typical two--dimensional fiber endowed with a
pseudo--Euclidean metric when $y^{3}=v=\varphi $ is the anisotropic
coordinate and $y^{4}=t$ is the time like coordinates. \ Such an exact
solution for the Einstein equations for the canonical d--connection is
described by a d--metric (\ref{solrot}) $\ $with coefficients of type (\ref%
{aux2a}).

\subsection{Nonholonomic transforms, Finsler variables and exact solutions
in (pseudo) Finsler gravity theories}

Finsler variables can be considered on any (pseudo) Riemannian manifold/
tangent bundle $\mathbf{V}$ if we prescribe a generating fundamental Finsler
function $F(x,y).$ This function induces canonical (Finsler) N-- and
d--connec\-tion structures, a class of N--adapted frames and a Sasaky type
d--metric $\mathbf{f}.$ By nonholonomic deforms, using corresponding
vierbein coefficients, such values can be related to an arbitrary d--metric
structure $\mathbf{g}$ on $\mathbf{V},$ in particular, to an exact solution $%
\mathbf{\mathring{g}}.$\footnote{%
In this work, the term exact solution refers to the Einstein equations for
the canonical d--connection $\widehat{\mathring{\mathbf{D}}}$ and/or the
Levi--Civita connection $\mathring{\nabla}.$ Usually, we state exactly what
kind of linear connections are used. In general, we can nonholonomically
transform a Finsler d--metric $\mathbf{f}$ into a d--metric $\mathbf{g},$
when at least one of such d--metrics is not a solution of any gravitational
field equations for the corresponding d--connections, and finally to deform
such a sequence of two transforms into an exact solution $\mathbf{\mathring{g%
}}$ but the this does not mean that all corresponding Ricci d--tensors, for
instance, will vanish in the case that one of such d--metric is a vacuum
solution. If we state for an N--adapted frame structure that $\ \mathbf{f=g=%
\mathring{g}},$ we argue that we introduce certain $\mathbf{f}$-- and/or $%
\mathbf{\mathring{g}}$--variables for a given (pseudo) Riemannian metric $%
\mathbf{g}$ and this will result in corresponding equalities of all
Ricci/Einstein d--tensors.}

Let us consider three sets of data:

\begin{itemize}
\item[a)] The values $\mathbf{f}_{\alpha \beta }=[\ f_{ij},\ f_{ab},\
^{c}N_{i}^{a}]$ (\ref{fsm}) for 
$\ ^{c}\mathbf{e}^{\alpha }=\ ^{c}\mathbf{e}_{\ \underline{\alpha }}^{\alpha
}e^{\underline{\alpha }}=(e^{i}=dx^{i},\ ^{c}\mathbf{e}^{a}=dy^{a}+\
^{c}N_{i}^{a}dx^{i}),$ 
with $\ ^{c}\mathbf{e}_{\ \underline{\alpha }}^{\alpha }=[\ \ ^{c}\mathbf{e}%
_{\ \underline{i}}^{i}=\delta _{\ \underline{i}}^{i},\ \ ^{c}\mathbf{e}_{\
\underline{a}}^{a}],$ $e^{\underline{\alpha }}=[dx^{\underline{i}},dy^{%
\underline{a}}],$ defines a (pseudo) Finsler space with canonical
N--connection $\ ^{c}N_{i}^{a}.$ We shall use the canonical d--connection $%
\widehat{\mathbf{\Gamma }}_{\ \alpha \beta }^{\gamma }$ (\ref{candcon})
computed for values $\mathbf{f}_{\alpha \beta }.$

\item[b)] The values $\mathbf{g}_{\alpha ^{\prime }\beta ^{\prime }}=[\
g_{i^{\prime }j^{\prime }},\ h_{a^{\prime }b^{\prime }},\ N_{i^{\prime
}}^{a^{\prime }}]$ (\ref{gpsm}) for 
$\ \mathbf{e}^{\alpha ^{\prime }}=\ \ \mathbf{e}_{\ \underline{\alpha }%
}^{\alpha ^{\prime }}e^{\underline{\alpha }}=(e^{i^{\prime }}=dx^{i^{\prime
}},\ \mathbf{e}^{a^{\prime }}=dy^{a^{\prime }}+N_{i^{\prime }}^{a^{\prime
}}dx^{i^{\prime }}),$ 
with $\ \mathbf{e}_{\ \underline{\alpha }}^{\alpha ^{\prime }}=[\ \ \mathbf{e%
}_{\ \underline{i}}^{i^{\prime }}=\delta _{\ \underline{i}}^{i^{\prime }},\
\ \mathbf{e}_{\ \underline{a}}^{a^{\prime }}],$ determines a general
(pseudo) Riemannian metric (for purposes of this work, the coefficients\ $%
N_{i^{\prime }}^{a^{\prime }}$ will be not general ones but taken to satisfy
some conditions of type $N_{i^{\prime }}^{a^{\prime }}=\mathring{N}%
_{i}^{a}). $

\item[c)] The coefficients $\mathbf{\mathring{g}}_{\alpha ^{\prime \prime
}\beta ^{\prime \prime }}=[\ \mathring{g}_{i^{\prime \prime }j^{\prime
\prime }},\ \mathring{h}_{a^{\prime \prime }b^{\prime \prime }},\ \mathring{N%
}_{i^{\prime \prime }}^{a^{\prime \prime }}]$ for 
$\ \mathbf{\mathring{e}}^{\alpha ^{\prime \prime }}=\ \ \mathbf{\mathring{e}}%
_{\ \underline{\alpha }}^{\alpha ^{\prime \prime }}e^{\underline{\alpha }%
}=(e^{i^{\prime \prime }}=dx^{i^{\prime \prime }},\ \mathbf{e}^{a^{\prime
\prime }}=dy^{a^{\prime \prime }}+\ \mathring{N}_{i^{\prime \prime
}}^{a^{\prime \prime }}dx^{i^{\prime \prime }}),$ 
with $\ \mathbf{e}_{\ \underline{\alpha }}^{\alpha ^{\prime \prime }}=[\ \
\mathbf{e}_{\ \underline{i}}^{i^{\prime \prime }}=\delta _{\ \underline{i}%
}^{i^{\prime \prime }},\ \ \mathbf{e}_{\ \underline{a}}^{a^{\prime \prime
}}],$ define a solution of nonholonomic Einstein equations for the canonical
d--connection, or its restriction to the case of the Levi--Civita
connection; for various classes of (non) holonomic Einstein spaces we can
chose $\mathbf{\mathring{g}}_{\alpha ^{\prime \prime }\beta ^{\prime \prime
}}$ to be defined by a d--metric (\ref{lambsol}) with any subsets of
coefficients subjected to respective conditions (\ref{coeflsoldc}), or (\ref%
{lcls}), for a nontrivial cosmological constant, and (\ref{vacdsolc}), or (%
\ref{vaclcsoc}), for vacuum configurations.
\end{itemize}

To model a (pseudo) Finsler geometry in general relativity we have to impose
the conditions $\ \mathbf{f=g=\mathring{g}.}$

\subsubsection{(Pseudo) Riemannian metrics in Finsler variables}

By frame transforms any data of type a) can be equivalently expressed as
data of type b) and inversely. For $\mathbf{f}_{\alpha \beta }=\ \mathbf{e}%
_{\ \alpha }^{\alpha ^{\prime }}\ \mathbf{e}_{\ \beta }^{\beta ^{\prime }}%
\mathbf{g}_{\alpha ^{\prime }\beta ^{\prime }},$ we write explicit
parametrizations
\begin{eqnarray}
\ f_{ij} &=&e_{\ i}^{i^{\prime }}e_{\ j}^{j^{\prime }}g_{i^{\prime
}j^{\prime }}\mbox{\ and  \ }~\ f_{ab}=e_{\ a}^{a^{\prime }}e_{\
b}^{b^{\prime }}\ g_{a^{\prime }b^{\prime }},  \label{auxeq1} \\
\ N_{i^{\prime }}^{a^{\prime }} &=&e_{i^{\prime }}^{\ i}e_{\ a}^{a^{\prime
}}\ ^{c}N_{i}^{a},\mbox{\ or \ }\ ^{c}N_{i}^{a}=e_{i}^{\ i^{\prime }}e_{\
a^{\prime }}^{a}\ N_{i^{\prime }}^{a^{\prime }},  \label{auxeq2}
\end{eqnarray}
were, for instance, $e_{a^{\prime }\ }^{\ a}$ is inverse to $e_{\
a}^{a^{\prime }}.$

For simplicity, we chose $g_{i^{\prime }j^{\prime }}=diag[g_{1^{\prime
}},g_{2^{\prime }}],$ $h_{a^{\prime }b^{\prime }}=diag[h_{3^{\prime
}},h_{4^{\prime }}]$ and $N_{i^{\prime }}^{a^{\prime }}=\left( N_{i^{\prime
}}^{3^{\prime }}=w_{i^{\prime }},N_{i^{\prime }}^{4^{\prime }}=n_{i^{\prime
}}\right) .$ The (pseudo) Finsler data $\ f_{ij},$ $\ f_{ab}$ and $\
^{c}N_{i}^{a}$ $=\left( \ ^{c}N_{i}^{3}=\ ^{c}w_{i},\ ^{c}N_{i}^{4}=\
^{c}n_{i}\right) $ are with diagonal matrices, $\ f_{ij}=diag[f_{1},f_{2}]$
and $\ f_{ab}=diag[f_{3},f_{4}],$ if the generating function is of type $F=\
^{3}F(x^{1},x^{2},v)$ $+\ ^{4}F(x^{1},x^{2},y)$ for some homogeneous
(respectively, on $y^{3}=v$ and $y^{4}=y)$ functions $\ ^{3}F$ and $\ ^{4}F.$%
\footnote{%
Of course, we can work with arbitrary generating functions $%
F(x^{1},x^{2},v,y)$ but this will result in off--diagonal (pseudo) Finsler
metrics in N--adapted bases, which would request a more cumbersome matrix
calculus.} For a diagonal representation with $e_{\ 4}^{3^{\prime }}=e_{\
3}^{4^{\prime }}=e_{\ 1}^{2^{\prime }}=e_{\ 2}^{1^{\prime }}=0,$ we satisfy
the conditions (\ref{auxeq1}) if%
\begin{equation}
e_{\ 1}^{1^{\prime }}=\pm \sqrt{\left| \frac{\ f_{1}}{g_{1^{\prime }}}%
\right| },e_{\ 2}^{2^{\prime }}=\pm \sqrt{\left| \frac{\ f_{2}}{g_{2^{\prime
}}}\right| }\ ,e_{\ 3}^{3^{\prime }}=\pm \sqrt{\left| \frac{\ f_{3}}{%
h_{3^{\prime }}}\right| },e_{\ 4}^{4^{\prime }}=\pm \sqrt{\left| \frac{\
f_{4}}{h_{4^{\prime }}}\right| }.  \label{aux21a}
\end{equation}

For any fixed values $\ f_{i},\ f_{a}$ and $\ ^{c}w_{i},^{c}n_{i}$ and given
$g_{i^{\prime }}$ and $h_{a^{\prime }},$ we can compute $w_{i^{\prime }}$
and $\ n_{i^{\prime }}$ as
\begin{eqnarray}
w_{1^{\prime }} &=&\pm \sqrt{\left| \frac{g_{1^{\prime }}\ f_{3}}{%
h_{3^{\prime }}\ f_{1}}\right| }\ ^{c}w_{1},w_{2^{\prime }}=\pm \sqrt{\left|
\frac{g_{2^{\prime }}\ f_{3}}{h_{3^{\prime }}\ f_{2}}\right| }\ ^{c}w_{2},
\label{aux23} \\
n_{1^{\prime }} &=&\pm \sqrt{\left| \frac{g_{1^{\prime }}\ f_{4}}{%
h_{4^{\prime }}\ f_{1}}\right| }\ ^{c}n_{1},n_{2^{\prime }}=\pm \sqrt{\left|
\frac{g_{2^{\prime }}\ f_{4}}{h_{4^{\prime }}\ f_{2}}\right| }\ ^{c}n_{2}
\notag
\end{eqnarray}%
solving the equations (\ref{auxeq2}).

\subsubsection{Anti--diagonal frame transforms and exact solutions}

It is also possible to define frame transforms relating data of type b) to
some data of type c) and inversely. In this case, the vierbein matrices
should be taken to be anti--diagonal in order to keep in mind the
possibility to relate data c) with some (pseudo) Finsler ones of type a).

Let us consider $\ \mathbf{\mathring{g}}_{\alpha ^{\prime \prime }\beta
^{\prime \prime }}=\ \mathbf{\mathring{e}}_{\ \alpha ^{\prime \prime
}}^{\alpha ^{\prime }}\ \mathbf{\mathring{e}}_{\ \beta ^{\prime \prime
}}^{\beta ^{\prime }}\mathbf{g}_{\alpha ^{\prime }\beta ^{\prime }}$
parametrized in the form%
\begin{equation}
\mathring{g}_{i^{\prime \prime }j^{\prime \prime }}=g_{i^{\prime }j^{\prime
}}\mathbf{\mathring{e}}_{\ i^{\prime \prime }}^{i^{\prime }}\mathbf{%
\mathring{e}}_{\ j^{\prime \prime }}^{j^{\prime }}+h_{a^{\prime }b^{\prime }}%
\mathbf{\mathring{e}}_{\ i^{\prime \prime }}^{a^{\prime }}\mathbf{\mathring{e%
}}_{\ j^{\prime \prime }}^{b^{\prime }},\ \mathring{h}_{a^{\prime \prime
}b^{\prime \prime }}=g_{i^{\prime }j^{\prime }}\mathbf{\mathring{e}}_{\
a^{\prime \prime }}^{i^{\prime }}\mathbf{\mathring{e}}_{\ b^{\prime \prime
}}^{j^{\prime }}+h_{a^{\prime }b^{\prime }}\mathbf{\mathring{e}}_{\
a^{\prime \prime }}^{a^{\prime }}\mathbf{\mathring{e}}_{\ b^{\prime \prime
}}^{b^{\prime }},  \label{secmap}
\end{equation}%
for an exact solution of Einstein equations determined by data $\mathbf{%
\mathring{g}}_{\alpha \beta }=[\mathring{g}_{i},\mathring{h}_{a},$ $%
\mathring{N}_{i}^{a}]$ in a N--elongated base $\mathbf{\mathring{e}}^{\alpha
}=(dx^{i},\mathbf{\mathring{e}}^{a}=dy^{a}+\mathring{N}_{i}^{a}dx^{i})$. For
$\mathbf{\mathring{e}}_{\ i^{\prime \prime }}^{i^{\prime }}=\delta _{\
i^{\prime \prime }}^{i^{\prime }},\mathbf{\mathring{e}}_{\ a^{\prime \prime
}}^{a^{\prime }}=\delta _{\ a^{\prime \prime }}^{a^{\prime }},$ we write (%
\ref{secmap}) as
\begin{equation}
\mathring{g}_{i^{\prime \prime }}=g_{i^{\prime \prime }}+h_{a^{\prime
}}\left( \mathbf{\mathring{e}}_{\ i^{\prime \prime }}^{a^{\prime }}\right)
^{2},~\mathring{h}_{a^{\prime \prime }}=g_{i^{\prime }}\left( \mathbf{%
\mathring{e}}_{\ a^{\prime \prime }}^{i^{\prime }}\right) ^{2}+h_{a^{\prime
\prime }},  \label{aux32a}
\end{equation}%
i.e. four equations for eight unknown variables $\mathbf{\mathring{e}}_{\
i^{\prime \prime }}^{a^{\prime }}$ and $\mathbf{\mathring{e}}_{\ a^{\prime
\prime }}^{i^{\prime }},$ and $\ \mathring{N}_{i^{\prime \prime
}}^{a^{\prime \prime }}=\mathbf{\mathring{e}}_{i^{\prime \prime }}^{\
i^{\prime }}\ \mathbf{\mathring{e}}_{\ a^{\prime }}^{a^{\prime \prime }}\
N_{i^{\prime }}^{a^{\prime }}=N_{i^{\prime \prime }}^{a^{\prime \prime }}.$
For instance, we can solve the algebraic system (\ref{aux32a}):
$\mathbf{\mathring{e}}_{\ 1^{\prime \prime }}^{3^{\prime }} = \pm \sqrt{%
\left| \left( \mathring{g}_{1^{\prime \prime }}-g_{1^{\prime \prime
}}\right) /h_{3^{\prime }}\right| },\mathbf{\mathring{e}}_{\ 2^{\prime
\prime }}^{3^{\prime }}=0,\mathbf{\mathring{e}}_{\ i^{\prime \prime
}}^{4^{\prime }}=0, \mathbf{\mathring{e}}_{\ a^{\prime \prime }}^{1^{\prime
}} =0, \mathbf{\mathring{e}}_{\ 3^{\prime \prime }}^{2^{\prime }}=0, \mathbf{%
\mathring{e}}_{\ 4^{\prime \prime }}^{2^{\prime }}=\pm \sqrt{\left| \left(
\mathring{h}_{4^{\prime \prime }}-h_{4^{\prime \prime }}\right)
/g_{2^{\prime }}\right| },$ for certain nontrivial values of metric
coefficients.

Using (\ref{aux23}), with $\mathring{N}_{i^{\prime \prime }}^{a^{\prime
\prime }}=N_{i^{\prime \prime }}^{a^{\prime \prime }\prime },$ we get
\begin{eqnarray}
\mathring{w}_{1^{\prime }} &=&\pm \sqrt{\left| \frac{g_{1^{\prime }}\ f_{3}}{%
h_{3^{\prime }}\ f_{1}}\right| }\ ^{c}w_{1},\ \mathring{w}_{2^{\prime }}=\pm
\sqrt{\left| \frac{g_{2^{\prime }}\ f_{3}}{h_{3^{\prime }}\ f_{2}}\right| }\
^{c}w_{2},  \label{aux33} \\
\mathring{n}_{1^{\prime }} &=&\pm \sqrt{\left| \frac{g_{1^{\prime }}\ f_{4}}{%
h_{4^{\prime }}\ f_{1}}\right| }\ ^{c}n_{1},\ \mathring{n}_{2^{\prime }}=\pm
\sqrt{\left| \frac{g_{2^{\prime }}\ f_{4}}{h_{4^{\prime }}\ f_{2}}\right| }\
^{c}n_{2}.  \notag
\end{eqnarray}%
From these formulas, we compute $g_{i^{\prime }},h_{a^{\prime }},$ when $%
g_{i^{\prime }}=\delta _{i^{\prime }}^{i^{\prime \prime }}g_{i^{\prime
\prime }},h_{a^{\prime }}=\delta _{a^{\prime }}^{a^{\prime \prime
}}h_{a^{\prime \prime }},\mathring{w}_{i^{\prime }}$ $=\delta _{i^{\prime
}}^{i^{\prime \prime }}\mathring{w}_{i^{\prime \prime }},\mathring{n}%
_{i^{\prime }}=\delta _{i^{\prime }}^{i^{\prime \prime }}\mathring{n}%
_{i^{\prime \prime }}.$ Introducing $g_{i^{\prime }},h_{a^{\prime }}$ into (%
\ref{aux32a}) for given $\mathring{g}_{i^{\prime \prime }},\mathring{h}%
_{a^{\prime \prime }},$ we can determine four values from eight ones, $%
\mathbf{\mathring{e}}_{\ i^{\prime \prime }}^{a^{\prime }}$ and $\mathbf{%
\mathring{e}}_{\ a^{\prime \prime }}^{i^{\prime }}.$

\subsubsection{Nonholonomic Einstein spaces and Finsler variables}

We summarize the main steps which allows us to transform a (pseudo) Finsler
d--metric into a general (pseudo) Riemannian one and then to relate both
such d--metrics to an exact solution of the Einstein equations. Of course,
such geometric/physical models became equivalent if they are performed for
the same canonical d--connection and/or Levi--Civita connection.

\begin{enumerate}
\item Let consider a solution for (non)holonomic Einstein spaces:
\begin{eqnarray*}
\mathbf{\mathring{g}} &=&\mathring{g}_{i}dx^{i}\otimes dx^{i}+\mathring{h}%
_{a}(dy^{a}+\mathring{N}_{j}^{a}dx^{j})\otimes (dy^{a}+\mathring{N}%
_{i}^{a}dx^{i}) \\
&=&\mathring{g}_{i}e^{i}\otimes e^{i}+\mathring{h}_{a}\mathbf{\mathring{e}}%
^{a}\otimes \mathbf{\mathring{e}}^{a}=\mathring{g}_{i^{\prime \prime
}j^{\prime \prime }}e^{i^{\prime \prime }}\otimes e^{j^{\prime \prime }}+%
\mathring{h}_{a^{\prime \prime }b^{\prime \prime }}\mathbf{\mathring{e}}%
^{a^{\prime \prime }}\otimes \mathbf{\mathring{e}}^{b^{\prime \prime }}
\end{eqnarray*}%
related to an arbitrary (pseudo) Riemannian metric transforms (\ref{secmap}).

\item We chose on $\mathbf{V}$ a fundamental (pseudo) Finsler function $F=\
^{3}F(x^{i},v)$ $+\ ^{4}F(x^{i},y)$ inducing canonically a d--metric of type
\begin{eqnarray*}
\ \mathbf{f} &=&\ f_{i}dx^{i}\otimes dx^{i}+\ f_{a}(dy^{a}+\
^{c}N_{j}^{a}dx^{j})\otimes (dy^{a}+\ ^{c}N_{i}^{a}dx^{i}), \\
&=&\ f_{i}e^{i}\otimes e^{i}+\ f_{a}\ ^{c}\mathbf{e}^{a}\otimes \ ^{c}%
\mathbf{e}^{a}
\end{eqnarray*}%
determined by data $\ \mathbf{f}_{\alpha \beta }=\left[ \ f_{i},\ f_{a},\
^{c}N_{j}^{a}\right] $ in a canonical N--elongated base $\ ^{c}\mathbf{e}%
^{\alpha }=(dx^{i},\ ^{c}\mathbf{e}^{a}=dy^{a}+\ ^{c}N_{i}^{a}dx^{i}).$

\item From formulas (\ref{aux33}) with $N_{i^{\prime }}^{a^{\prime }}=%
\mathring{N}_{i^{\prime }}^{a^{\prime }}$ and $\mathbf{e}^{\alpha ^{\prime
}}=\mathbf{\mathring{e}}^{\alpha ^{\prime }},$ we obtain
\begin{equation*}
g_{i^{\prime }}=\ f_{i^{\prime }}\left( \frac{\mathring{w}_{i^{\prime }}}{\
^{c}w_{i^{\prime }}}\right) ^{2}\frac{h_{3^{\prime }}}{\ f_{3^{\prime }}},\
g_{i^{\prime }}=\ f_{i^{\prime }}\left( \frac{\mathring{n}_{i^{\prime }}}{\
^{c}n_{i^{\prime }}}\right) ^{2}\frac{h_{4^{\prime }}}{\ f_{4^{\prime }}}.
\end{equation*}%
Both formulas are compatible if $\mathring{w}_{i^{\prime }}$ and $\mathring{n%
}_{i^{\prime }}$ are constrained (this is possible if we chose (\ref%
{coeflsoldc}) and (\ref{vacdsolc})) to satisfy the conditions $\Theta
_{1^{\prime }}=\Theta _{2^{\prime }}=\Theta ,$ where $\Theta _{i^{\prime
}}=\left( \frac{\mathring{w}_{i^{\prime }}}{\ ^{c}w_{i^{\prime }}}\right)
^{2}\left( \frac{\mathring{n}_{i^{\prime }}}{\ ^{c}n_{i^{\prime }}}\right)
^{2},$ \ and $\Theta =\left( \frac{\mathring{w}_{1^{\prime }}}{\
^{c}w_{1^{\prime }}}\right) ^{2}\left( \frac{\mathring{n}_{1^{\prime }}}{\
^{c}n_{1^{\prime }}}\right) ^{2}=\left( \frac{\mathring{w}_{2^{\prime }}}{\
^{c}w_{2^{\prime }}}\right) ^{2}\left( \frac{\mathring{n}_{2^{\prime }}}{\
^{c}n_{2^{\prime }}}\right) ^{2}.$ Having computed $\Theta ,$ we define $%
g_{i^{\prime }}=\left(\frac{\mathring{w}_{i^{\prime }}}{\ ^{c}w_{i^{\prime }}%
}\right) ^{2}\frac{\ f_{i^{\prime }}}{f_{3^{\prime }}}$ and $h_{3^{\prime
}}=h_{4^{\prime }}\Theta ,$ where (in this case) there is not summing on
indices. So, we constructed the data $g_{i^{\prime }},h_{a^{\prime }}$ and $%
w_{i^{\prime }},n_{j^{\prime }}.$

\item We can construct $\mathbf{\mathring{e}}_{\ i^{\prime \prime
}}^{a^{\prime }}$ and $\mathbf{\mathring{e}}_{\ a^{\prime \prime
}}^{i^{\prime }}$ as any nontrivial solutions of
\begin{equation*}
\mathring{g}_{i^{\prime \prime }}=g_{i^{\prime \prime }}+h_{a^{\prime
}}\left( \mathbf{\mathring{e}}_{\ i^{\prime \prime }}^{a^{\prime }}\right)
^{2},\ \mathring{h}_{a^{\prime \prime }}=g_{i^{\prime }}\left( \mathbf{%
\mathring{e}}_{\ a^{\prime \prime }}^{i^{\prime }}\right) ^{2}+h_{a^{\prime
\prime }},\ \mathring{N}_{i^{\prime \prime }}^{a^{\prime \prime
}}=N_{i^{\prime \prime }}^{a^{\prime \prime }}.
\end{equation*}%
For instance, we take 
$\mathbf{\mathring{e}}_{\ 1^{\prime \prime }}^{3^{\prime }} =\pm \sqrt{%
\left| \left( \mathring{g}_{1^{\prime \prime }}-g_{1^{\prime \prime
}}\right) /h_{3^{\prime }}\right| },\mathbf{\mathring{e}}_{\ 2^{\prime
\prime }}^{3^{\prime }}=0,\mathbf{\mathring{e}}_{\ i^{\prime \prime
}}^{4^{\prime }}=0,$ $\mathbf{\mathring{e}}_{\ a^{\prime \prime
}}^{1^{\prime }} =0,\mathbf{\mathring{e}}_{\ 3^{\prime \prime }}^{2^{\prime
}}=0,\mathbf{\mathring{e}}_{\ 4^{\prime \prime }}^{2^{\prime }}=\pm \sqrt{%
\left| \left( \mathring{h}_{4^{\prime \prime }}-h_{4^{\prime \prime
}}\right) /g_{2^{\prime }}\right| }$ 
and finally compute
\begin{equation*}
e_{\ 1}^{1^{\prime }}=\pm \sqrt{\left| \frac{\ f_{1}}{g_{1^{\prime }}}%
\right| },\ e_{\ 2}^{2^{\prime }}=\pm \sqrt{\left| \frac{\ f_{2}}{%
g_{2^{\prime }}}\right| },\ e_{\ 3}^{3^{\prime }}=\pm \sqrt{\left| \frac{\
f_{3}}{h_{3^{\prime }}}\right| },\ e_{\ 4}^{4^{\prime }}=\pm \sqrt{\left|
\frac{\ f_{4}}{h_{4^{\prime }}}\right| }.
\end{equation*}
\end{enumerate}

We note that we defined a sequence of two nonholonomic deformations from $\
\mathbf{f}$ to $\mathbf{\mathring{g}}$ and inversely. The above geometric
constructions are outlined in Table \ref{tabl1}.
\begin{table}[tbp]
{\scriptsize
\begin{tabular}{|c|c|c|c|c|}
\hline\hline
$\mathbf{f}$ & $\leftrightarrow $ & $\mathbf{g}$ & $\leftrightarrow $ & $%
\mathring{\mathbf{g}}$ \\
$\shortparallel $ &  & $\shortparallel $ &  & $\shortparallel $ \\
$\{\mathbf{f}_{\alpha \beta }\}$ &  & $\{\mathbf{f}_{\alpha \beta }=\mathbf{e%
}_{\ \alpha }^{\alpha ^{\prime }}\ \mathbf{e}_{\ \beta }^{\beta ^{\prime }}%
\mathbf{g}_{\alpha ^{\prime }\beta ^{\prime }}\}$ &  & $\{\mathbf{g}_{\alpha
^{\prime }\beta ^{\prime }}=\ \mathbf{\mathring{e}}_{\ \alpha ^{\prime
}}^{\alpha ^{\prime \prime }}\ \mathbf{\mathring{e}}_{\ \beta ^{\prime
}}^{\beta ^{\prime \prime }}\mathbf{\mathring{g}}_{\alpha ^{\prime \prime
}\beta ^{\prime \prime }}\}$ \\ \hline\hline
&  & $\mathbf{e}_{\ \alpha }^{\alpha ^{\prime }}=[e_{\ i}^{i^{\prime }},%
\mathbf{\mathring{e}}_{\ a}^{a^{\prime }}]$ &  & ${\left.
\begin{array}{c}
\ \mathbf{\mathring{e}}_{\ i^{\prime \prime }}^{a^{\prime }}=\left[
\begin{array}{c}
\mathbf{\mathring{e}}_{\ 1^{\prime \prime }}^{3^{\prime }};\mathbf{\mathring{%
e}}_{\ a^{\prime \prime }}^{a^{\prime }}=\delta _{\ a^{\prime \prime
}}^{a^{\prime }} \\
\mathbf{\mathring{e}}_{\ i^{\prime \prime }}^{4^{\prime }}=\mathbf{\mathring{%
e}}_{\ 2^{\prime \prime }}^{3^{\prime }}=0%
\end{array}%
\right] \\
\ \mathbf{\mathring{e}}_{\ a^{\prime \prime }}^{i^{\prime }}=\left[
\begin{array}{c}
\mathbf{\mathring{e}}_{\ i^{\prime \prime }}^{i^{\prime }}=\delta _{\
i^{\prime \prime }}^{i^{\prime }};\mathbf{\mathring{e}}_{\ 4^{\prime \prime
}}^{2^{\prime }} \\
\mathbf{\mathring{e}}_{\ a^{\prime \prime }}^{1^{\prime }}=\mathbf{\mathring{%
e}}_{\ 3^{\prime \prime }}^{2^{\prime }}=0%
\end{array}%
\right] \\
\mathbf{\mathring{e}}_{\ \alpha ^{\prime \prime }}^{\alpha ^{\prime }}=[%
\mathbf{\mathring{e}}_{\ i^{\prime \prime }}^{a^{\prime }},\mathbf{\mathring{%
e}}_{\ a^{\prime \prime }}^{i^{\prime }}]%
\end{array}%
\right. }$ \\ \hline
&  &  &  &  \\
$\left.
\begin{array}{c}
\begin{array}{c}
\ f_{ij}=diag\{f_{i}\} \\
\ f_{ab}=diag\{f_{a}\}%
\end{array}
\\
\ ^{c}N_{i}^{a}=\left\{
\begin{array}{c}
\ ^{c}w_{i} \\
\ ^{c}n_{i}%
\end{array}%
\right. \\
\ ^{c}\mathbf{e}^{\alpha }=(dx^{i},\ ^{c}\mathbf{e}^{a})%
\end{array}%
\right. $ &  & $\left.
\begin{array}{c}
\begin{array}{c}
\ g_{i^{\prime }j^{\prime }}=diag\{g_{i^{\prime }}\} \\
\ h_{a^{\prime }b^{\prime }}=diag\{\ h_{a^{\prime }}\}%
\end{array}
\\
\ N_{i^{\prime }}^{a^{\prime }}=\ \mathring{N}_{i^{\prime }}^{a^{\prime }}
\\
\ \mathbf{e}^{\alpha ^{\prime }}=(dx^{i},\ \mathbf{\mathring{e}}^{a^{\prime
}})%
\end{array}%
\right. $ &  & $\left.
\begin{array}{c}
\begin{array}{c}
\ \mathring{g}_{i^{\prime \prime }j^{\prime \prime }}=diag\{\ \mathring{g}%
_{i^{\prime \prime }}\} \\
\ \mathring{h}_{a^{\prime \prime }b^{\prime \prime }}=diag\{\ \mathring{h}%
_{a^{\prime \prime }}\}%
\end{array}
\\
\ \mathring{N}_{i^{\prime \prime }}^{a^{\prime \prime }}=\left\{
\begin{array}{c}
\ \mathring{w}_{i^{\prime \prime }} \\
\mathring{n}_{i^{\prime \prime }}%
\end{array}%
\right. \\
\mathbf{\mathring{e}}^{\alpha ^{\prime \prime }}=(\mathbf{\mathring{e}}%
^{i^{\prime \prime }},\ \mathbf{\mathring{e}}^{a^{\prime \prime }})%
\end{array}%
\right. $ \\ \hline\hline
\end{tabular}
}
\caption{Nonholonomic deformations of (pseudo) Finsler metrics into (pseudo)
Riemannian/ Einstein ones.}
\label{tabl1}
\end{table}

The goal of this section was to prove that for any model of (pseudo) Finsler
gravity induced by a generating function of type $F=\ ^{3}F(\xi ,\vartheta
,\varphi )$ $+\ ^{4}F(\xi ,\vartheta ,\varphi )$ there are exact solutions
with rotoid symmetry, of type (\ref{lambsol}), for Einstein equations with
nontrivial \ cosmological constant. In the limit $\varepsilon \rightarrow 0$
for $\mathbf{\mathring{g}},$ the elaborated scheme of two nonholonomic
transforms allows us to rewrite the Schwarzschild solution as a (pseudo)
Finsler metric $\ \mathbf{f}(x,y)\mathbf{.}$ Haven chosen to define our
gravity theory on a N--anholonomic manifold, we say that the the
Schwarzschild spacetime is parametrized in (nonholonomic) Finsler variables.

Any construction on nonholonomic (pseudo) Riemannian spaces can be similarly
performed for Finsler gravity theories on tangent bundles. In such a case,
the variables $y^{a}$ must be interpreted as ''velocities'' and the
fundamental geometric objects (the metric and N-- and d--connections) will
depend on such tangent vectors components. A natural Schwarzschild like
generalization of $\mathbf{\mathring{g}}$ would be to chose a d--metric $%
\mathbf{g}_{\alpha ^{\prime }\beta ^{\prime }}=[\ g_{i^{\prime }j^{\prime
}},\ h_{a^{\prime }b^{\prime }},\ N_{i^{\prime }}^{a^{\prime }}]$ (\ref{gpsm}%
) included in a scheme $\mathbf{f\leftrightarrow g\leftrightarrow \mathring{g%
}}$ when the canonical d--connection $\widehat{\mathbf{\Gamma }}_{\ \alpha
\beta }^{\gamma }$ (\ref{candcon}) is for a solution of nonholonomic vacuum
Einstein equations with $h_{4}$ and $h_{3}$ defined respectively by $b^{2}=q$
and $\left( b^{\ast }\right) ^{2}=\left[ (\sqrt{|q|})^{\ast }\right] ^{2}$
introduced in (\ref{rotoidm}) but with general N--connection coefficients (%
\ref{vacdsolc}). Such configurations seem to be stable and define
(nonholonomic) black hole objects in (pseudo) Finsler gravity (we have to
chose correspondingly the integration functions $^{1}n_{i}(\xi ,\vartheta
),\ ^{2}n_{i}(\xi ,\vartheta )$ and the coefficients $w_{1}(\xi ,\vartheta
,\varphi )$ and adapt the proof for ''black ellipsoids'' from \cite%
{vbe1,vbe2,vncg}).

\section{Discussion and Conclusions}

In this paper, we have analyzed the problem of how black hole solutions can
be constructed in Finsler gravity theories and if such geometric objects may
have relations to black holes in general relativity and generalizations. To
the best of our knowledge, such questions have not yet been addressed in the
literature.

There are also two other important motivations to study possible black hole
structures and their nonholonomic deformations on (pseudo) Riemannian
spacetimes and gravity models on tangent bundles:

\begin{enumerate}
\item To analyze the physical consequences of nonholonomic frame constraints
on the dynamics of gravitational fields induced by spacetime nonholonomic
gravitational distributions and/or nonlinear self--polarizations of
gravitational fields.

\item To provide additional arguments on viability of Finsler like gravity
theories. If such models admit black hole objects which are non--trivially
related to those in general relativity, this may help a better understanding
of stationary gravitational configurations and their modelling by
Lagrange--Finseler geometries.
\end{enumerate}

Our constructions have completed a qualitative understanding of a class of
exact solutions in the Einstein and Finsler gravity theories which for
certain small values of parametric nonholonomic deformations contain stable
ellipsoid configurations and new classes of black hole objects. There were
encompassed all possible values of cosmological constant for solutions with
generic off--diagonal metrics and two classes of linear connections (the
canonical distinguished connection and the Levi--Civita one). The solutions
were generated following the anholonomic frame method (see reviews of such
geometric methods and results in Refs. \cite{ijgmmp,vrflg,vsgg}). Certain
features of these solutions are shared, while others differ or can be
modelled in certain limits of a small parameter and for some types of
generating/integrating functions. For instance, we positively get black hole
solutions with ellipsoidal symmetry for certain small values of
eccentricity, but dependence on cosmological constant plays not a smooth
character because of nonlinear interactions and nonholonomic constraints. At
the most basic level, we have to introduce locally anisotropic polarizations
of masses in order to get self--consistent and stable gravitational
configurations.

In the context of interactions of Finsler black hole solutions with
nonlinear waves (we have chosen the example of solitonic waves), our
geometric method allows us to include them both as generic off--diagonal
terms and/or in the so--called ''vertical'' part of the metrics as small and
not small deformations of original black holes spacetimes. We can consider
various types of asymptotic conditions and nonholonomic constraints and
define stationary stable configurations.

In another context of nonholonomic Einstein spacetimes modelled on (pseudo)
Riemannian/ Finsler manifolds/bundles, we found that the same classes of
black hole solutions, rotoids and/or solitons can be derived in all metric
compatible gravity theories. The results of this paper have lead to an
overview of the main qualitative features common to static solutions in
general relativity and their stationary modifications for Finsler like
theories. These results confirm and extend our results and knowledge on the
existing/possible generalizations for black hole solutions in string/brane
models, noncommutative gravity, nonholonomic Ricci flow theory etc, see
examples and discussions in Refs. \cite{vs5dbh,vs5dbh1,vrf1,vrf2,vrf5}.

\vskip5pt

\textbf{Acknowledgement: } This research is partially supported by the
Program IDEI, PN-II-ID-PCE-2011-3-0256. Author is grateful to referees for
hard work and very important requests and suggestions to introduce new
references on gravity models on contangent bundles and a Glossary of terms.

\appendix

\setcounter{equation}{0} \renewcommand{\theequation}
{A.\arabic{equation}} \setcounter{subsection}{0}
\renewcommand{\thesubsection}
{A.\arabic{subsection}}

\section{Glossary of Terms and Comments}

We provide a list with basic terms, concepts and terms using "Italic"
characters in (generalized) Finsler and Lagrange geometry and nonholonomic
(pseudo) Riemannian spaces. Usually, such models are elaborated on a tangent
bundle $TM$ of a manifold $M,$ $\dim M=n\geq 2,$ or (in Cartan and Hamilton
geometry) on dual, i.e. co--tangent, bundle $T^{\ast }M.$ Indices $i,j,...$
are used for coordinates on $M$ and indices $a,b,...$ are used for fiber
like coordinates. The Greek indices $\alpha ,\beta ,...$ are for coordinates
$u^{\alpha }=(x^{i},y^{a}),$ or $u=(x,y),$ and geometric objects on total
space $TM.$ References \cite{vacaru2012,vcritic,vrflg,vsgg} provide surveys,
critical remarks and details on physical models for generalized Finsler
geometry and applications in modern physics.

\begin{enumerate}
\item A \textit{Finsler fundamental (generating) function} $F(x,y),$ when $%
F(x,\lambda y)=|\lambda |F(x,y),$ for any nonzero $\lambda \in \mathbb{R},$
is defined on $\widetilde{TM}$ (which is $TM$ without null--sections on $M$;
some additional assumptions on $F$ are imposed in different geometric models
of Finsler geometry, see details in section \ref{ssfg}). The nonlinear
quadratic element
 $ds^{2}=F^{2}(x,y)$ 
transform into the well--known Riemannian one,
 $ds^{2}=g_{ij}(x)dx^{i}dx^{k}$, 
if $F^{2}$ is a quadratic function on $y^{i}\sim dx^{i}.$ In literature on
geometry and physics, there is a confusion between the terms "Finsler
metric" extending the concept of "Riemannian metric" $g_{ij}(x)$ on $M.$
Some authors call $F$ as the Finsler metric but other ones use this term for
the Hessian $\ f_{ab}=\frac{1}{2}\frac{\partial ^{2}F^{2}}{\partial
y^{a}\partial y^{b}},$ $\det |\ f_{ab}|\neq 0,$ see (\ref{efm}), or for a
metric defined by a Sasaki lift $\mathbf{f}_{\alpha \beta }$ (\ref{fsm}). To
avoid ambiguities is better to call  $F$ as the 
generating/fundamental Finsler function because different metric structure
can be constructed on $TM,$ $M$ and fibers using different maps for the same
geometric data $(M,F).$

\item We can elaborate "effective" \textit{mechanical models of Finsler
geometry} if $L=F^{2}$ is considered as a regular Lagrangian. Such models
can be related to "relativistic mechanics" if the condition of quadratic
positivity is dropped for $\ f_{ab}$ and there are considered
pseudo--Euclidean signatures for $g_{ij}(x)$ and $\mathbf{f}_{\alpha \beta
}(x,y).$ A (pseudo) Riemannian space $(M,g_{ij}(x))$ is characterized by a
unique Levi--Civita connection $\nabla =\{\Gamma _{jk}^{i}\}$ which is
torsionless and metric compatible. For such a connection, the equations for
"autoparallel transports" \ and geodesics (length extremals)
are equivalent. In general, it is not clear which conditions should be stated when the
equations for nonlinear geodesics (semi--sprays) would be equivalent to
certain "autoparallel" equations in Finsler space.

\item The \textit{main confusion} for non-experts on Finsler geometry is
that a metric $g_{ij}$ completely defines a (pseudo) Riemannian geometry but
only $F$ does not determine a complete model of (pseudo) Finsler geometry.
There are necessary additional assumptions (following certain
geometric/physical principles) on which types of connections have to be used
for such spaces. One of the approaches to definition of Finsler spaces is to
work with "semi--sprays" as certain variational curves and related
Euler--Lagrange equations for $L=F^{2}.$ In such a case, we can construct
\textit{the Cartan nonlinear connection} $\ ^{c}\mathbf{N}=\{\
^{c}N_{i}^{a}\}$, see formulas related to (\ref{clnc}), which is completely
defined by $F$ up to arbitrary frame/coordinate transforms.

\item In general, a \textit{nonlinear connection (N--connection)} can be
considered as nonholonomic (nonintegrable) distribution stating a general
splitting $\mathbf{N}:TTM=hTM\oplus vTM,$ see (\ref{whitney}). Such a
horizontal (h) -- vertical (v) decomposition is not obligatory defined
for a generating function $F.$

\item One of the most important geometric and physical property of a
N--connection $\ \mathbf{N}=\{N_{i}^{a}\},$ in a canonical or other forms,
is that it allows us to introduce the so--called \textit{N--adapted frames
and coframes} (\ref{ddif})
\begin{equation*}
\mathbf{e}_{\nu }=(\mathbf{e}_{i}=\partial _{i}-N_{i}^{a}\partial
_{a},e_{a}=\partial _{a}),\mathbf{e}^{\mu }=e^{i}=dx^{i},\mathbf{e}%
^{a}=dy^{a}+N_{i}^{a}dx^{i}.
\end{equation*}%
We can elaborate N--adapted differential/integral and variational calculus
with respect to such nonholonomic vielbeins (frames).

\item Via so--called Sasaki type lifts, using the Hessian $\ f_{ab}$ (\ref%
{efm}) and the canonical N--connection $\ ^{c}\mathbf{N,}$ a fundamental
Finsler function $F$ defines a so--called \textit{canonical distinguished
metric (d--metric) }structure $\mathbf{f}_{\alpha \beta }$ (\ref{fsm}) on $%
TM.$ Some authors prefer to consider other classes of lifts when the
homogeneity property is preserved but for Finsler like extensions of general
relativity with general covariance we can work with any $\mathbf{f}_{\alpha
\beta }$. This allows us to consider any convenient frames and systems of
coordinates which is very important for constructing exact solutions and
quantization.

\item The terms \textit{distinguished tensor (d--tensor), distinguished
vector (d--vector),} etc are used for geometric objects adapted to a
N--connection splitting (\ref{whitney}). We can use the data $(\mathbf{f}%
_{\alpha \beta },\ ^{c}N_{i}^{a})$ and construct on $TM$ a canonical
(pseudo) Riemannian model determined by a fundamental Finsler function $F$
and the Levi--Civita connection $\ ^{F}\nabla $ computed for $\mathbf{f}%
_{\alpha \beta }.$ But in such a "Finsler" geometry the constructions will
be not adapted the the N--connection splitting (\ref{whitney}). In Finsler
geometry, one works with with linear connections which preserve the
N--connection structure. The Levi--Civita connection does not have such a
property.

\item A \textit{distinguished connection (d--connection), }$\mathbf{D}%
=(hD,vD)=\{\mathbf{\Gamma }_{\ \alpha \beta }^{\gamma
}=(L_{jk}^{i},L_{bk}^{a},C_{jc}^{i},C_{bc}^{a})\},$ is a linear connection
preserving under parallel maps the $h$--$v$--decompositions of geometric
objects. In Finsler geometry, there are used various classes of
d--connections which a metric compatible, $\mathbf{Dg=0,}$ or not metric
compatible, $\mathbf{Dg\neq 0.}$ They are defined following certain
geometric principles

\item There is a so--called \textit{canonical d--connection} $\widehat{%
\mathbf{D}}=\{\widehat{\mathbf{\Gamma }}_{\ \alpha \beta }^{\gamma }\}$ (\ref%
{candcon}) which by definition is metric compatible, $\widehat{\mathbf{D}}%
\mathbf{g=0,}$ and with zero $h$- and $v$--torsion. The nontrivial $h$-$v$%
--components of such a d--torsion are completely determined by a d--metric $%
\mathbf{g.}$ Such a d--connection is very convenient for constructing
solutions of gravitational field equations in Einstein and Finsler gravity
theories, and modifications, see \cite%
{vrflg,vsgg,ijgmmp,vncg,vs5dbh,vs5dbh1,vgsol}.

\item Chronologically, the first Finsler type d--connection was introduced
by Cartan \cite{cart}, \textit{the Cartan d--connection, }$\mathbf{\tilde{D},%
}$ for certain identifications $\widetilde{L}_{jk}^{i}=\widetilde{L}%
_{bk}^{a},\widetilde{C}_{jc}^{i}=\tilde{C}_{bc}^{a}$ when all geometric
objects are generated by $F.$ This d--connection has a very important
property that it is compatible to the almost symplectic (Kaehler) and amost
complex structures constructed in a canonical form using $(\mathbf{f}%
_{\alpha \beta },\ ^{c}N_{i}^{a}).$ As a result, it is possible to quantize
such theories using the Fedosov and/or A--brane quantization methods \cite%
{vpla,vbrane,av}, also references therein.

\item If $\mathbf{g=f,}$ the coefficients of $\widehat{\mathbf{D}}$ and/or $%
\mathbf{\tilde{D}}$ are completely determined by $F$ and the corresponding $%
\mathbf{f}_{\alpha \beta }.$ In such cases, there are \textit{N--adapted
distortions to the Levi--Civita connection}, for instance, in the form $%
\Gamma _{\ \alpha \beta }^{\gamma }=\widehat{\mathbf{\Gamma }}_{\ \alpha
\beta }^{\gamma }+Z_{\ \alpha \beta }^{\gamma },$ see formulas (\ref{deflc})
and (\ref{deft}). We can work equivalently with data $(F:\mathbf{g=f,}%
\widehat{\mathbf{D}}),$ or $(F:\mathbf{g=f,\tilde{D}}),$ or $(F:\mathbf{%
g=f,\ ^{F}\nabla =\widehat{\mathbf{D}}+}\widehat{\mathbf{Z}})$ etc. The
corresponding torsion and curvature tensors can be constructed/computed in
standard form but for resepective d--connections.

\item Finsler like connections can be used on a (pseudo) Riemannian space, $%
V,$ and in "standard" general relativity theory (GR). This is possible for
any 2+2 splitting with corresponding fibred structure for conventional $h$-
and $v$--coordinates, $u^{\alpha }=(x^{i},y^{a}),$ or $u=(x,y),$ when $%
i,j,...=1,2$ and $a,b,...=3,4$. We can prescribe any formal Finsler like
nonholonomic distribution $\mathcal{F}$ on $V$ (analog of $F$ on $TM).$
Similarly to constructions on tangent bundles, we can define and compute $(%
\mathbf{f}_{\alpha \beta },\ ^{c}N_{i}^{a})$ and related $\widehat{\mathbf{D}%
}$ and $\mathbf{\tilde{D},}$ when the nonholonomically induced torsion is
completely determined by the metric structure. This is the main difference
from the well--known Cartan geometry when the torsion is an independent
tensor field). We conclude that \textit{any (pseudo) Riemann geometry and
the GR can be equivalently encoded in Finsler like variables (similarly, we
can re--write such models using spinor, twistor or other variables), and
certain equivalent almost symplectic ones. } Some variables are more
convenient, for instance, for constructing generic off--diagonal exact
solutions, other ones for analyzing interactions with fermions in N--adapted
form, or to re--define the constructions for a geometric method of
quantization.

\item \textit{Conclusions: } A model of Finsler geometry is completely
defined by a triple of data $(F:\mathbf{f,N,D})$ which is different from a
(pseudo) Riemannian geometry determined only by a metric field $%
\mathbf{g}$ (in GR, it is a solution of the Einstein equations). We need additional assumptions on Finsler types N-- and d--connection  and metric structures.

\item \textit{Final Remarks:}

\begin{itemize}
\item  There are different models of Finsler geometry. Various authors
consider Finsler d--connections which are not metric compatible (for
instance, the Chern and/or Berwald d--connections), or different definitions
for the Riemann and Ricci tensors, see critical remarks and discussions on
relation of such models to standard theories of physics in Refs. \cite%
{vcritic,vrflg,vsgg}.

\item  In Ref. \cite{vacaru2012}, we concluded that a self--consistent
axiomatics for a class of Einstein--Finsler theories constructed for the
d--connections $\widehat{\mathbf{D}}$ and/or $\mathbf{\tilde{D}}$ (used
instead of the Levi--Civita one $\nabla )$ can be formulated in a form
similar to GR. If we take the basic space $M$ as a Lorentz manifold, the
(co) tangent bundle constructions encode Finsler like generalizations of the
Einstein theory in certain unique canonical forms. Such theories can be
quantized following standard methods. They allow us to study certain models
with Lorentz violations, modified gravity effects etc. We can apply pure
geometric methods or elaborate N--adapted variational and generalized
Palatini approaches with N--elongated partial derivatives and differentials (%
\ref{ddif}). For such details, see \cite{vrflg,vsgg} and references therein.
\end{itemize}
\end{enumerate}

\end{document}